*Type of the Paper: Article*
# Complexity Analysis of Environmental Time Series

Holger Lange[1,*] and Michael Hauhs[2,*]

[1] Norwegian Institute of Bioeconomy Research (NIBIO), N-1433 Ås, Norway; holger.lange@nibio.no
[2] University of Bayreuth, Germany; michael.hauhs@uni-bayreuth.de
[*] Correspondence: holger.lange@nibio.no; Tel.: +47 90088460

**Abstract:** Small, forested catchments are prototypes of terrestrial ecosystems and have been studied in several disciplines of environmental sciences since several decades. Time series of water and matter fluxes and nutrient concentrations from these systems exhibit a bewildering diversity of spatiotemporal patterns, indicating the intricate nature of processes acting on a large range of time scales. Nonlinear dynamics is an obvious framework to investigate catchment time series. We analyze selected long-term data from three headwater catchments in the Bramke valley, Harz mountains, Lower Saxony in Germany at common biweekly resolution for the period 1991 to 2023. For every time series, we perform gap filling, detrending and removal of the annual cycle using Singular System Analysis (SSA), and then calculate metrics based on ordinal pattern statistics: the permutation entropy, permutation complexity and Fisher information, as well as their generalized versions (q-entropy and $\alpha$-entropy). Further, the position of each variable in Tarnopolski diagrams is displayed and compared to reference stochastic processes, like fractional Brownian motion, fractional Gaussian noise, and $\beta$ noise. Still another way of distinguishing deterministic chaos and structured noise, and quantifying the latter, is provided by the complexity from ordinal pattern positioned slopes (COPPS). We also construct Horizontal Visibility Graphs and estimate the exponent of the decay of the degree distribution. Taken together, the analyses create a characterization of the dynamics of these systems which can be scrutinized for universality, either across variables or between the three geographically very close catchments.

**Keywords:** time series; ordinal patterns; catchments; ecosystems; permutation entropy; permutation complexity; Fisher information; Tarnopolski diagrams; Horizontal Visibility Graphs

## 1. Introduction

Long-term monitoring of terrestrial ecosystems is a key activity producing insights into trends, pertinent oscillations, and system dynamics in general. It is the backbone of statements about changes in the environment on different time scales, whether these are natural phenomena (e.g. succession), related to human activities like land use change, or to climate change. The monitoring programs generate long-term time series, often spanning several decades, and Earth System Models (ESMs) are attempting to reproduce the observations assuming a set of processes and eventually try to predict them using scenarios like the Shared Socioeconomic Pathways (SSPs), or to classify them by the urgency of intervention, in the case of management related variables.

ESMs have to assume a set of processes acting within the system and across its boundaries and need to be parametrized prior to simulations. The set of parameters needed for model calibration is often beyond what monitoring can deliver, and assumptions have to be made for the values of unobserved (or even unobservable) parameters. This is often done in an inverse modelling approach, where minimizing a cost function describing the data-model discrepancy is used to estimate the parameters needed. This is the recipe of most process-based approaches in the environmental sciences.

The alternative, data-driven approach is not assuming given processes and requires fewer or even no parameters to be estimated. It starts with a set of observed data, often time series, and concludes on their multivariate spatiotemporal structure. This is the route we also follow in this article.

We aim at a thorough characterization of this spatiotemporal structure through a set of metrics obtained from methods from nonlinear dynamics. These metrics separate deterministic from stochastic parts or the time series, elucidate the stochastic properties of them and provide insights into their information content and complexity, thereby also indicating the efforts needed to successfully model them process-based. It is reasonable to assume that reproducing rather complex data might also require complex models, although there might be exceptions. The opposite is not necessarily true: there are simple processes generating complex data, as convincingly demonstrated by toy chaotic maps like the logistic map or the Rössler attractor, and so on. From the many approaches to investigate the complexity of time series [1], we focus mainly on those where Osvaldo Rosso had a leading role or made significant contributions to [2-10], and predominantly methods utilizing ordinal pattern statistics.

Using a set of variables across several locations allows to investigate the classic question whether the dynamics of a given variable (here: ion concentrations in water solutions) is universal for that variable, governed by the same processes at different locations (spatial universality), or, alternatively, the location (point of measurement) determines the dynamics (temporal universality), i.e. we observe similar dynamics within the given ecosystem, but the same variables at different locations show diverging dynamics. The modelling framework suitable in each case might be rather different. Classifying the data set by appropriate metrics supports decisions about the most suitable modelling approach. It is, however, notoriously difficult to reproduce all of the complexity metrics with any process-based model.

Two typical modelling approaches prominent in environmental sciences, in particular in forest science and in hydrology in our case, can be described as follows: firstly, physical transport models based on a dynamic system approach (the Richards equations solved with appropriate boundary conditions [11]) and secondly forest growth models (e.g. yield tables [12] or growth simulators [13]) based on regional growth histories of the same species and treatment. The first one is focussing on abiotic aspects of the system and keeps the acting organisms at an abstract level; the second one considers the growing trees and keeps the physicochemical aspects at a rather simplified level. Of course, in between these cartoon representations there is a continuum of hybrid model classes, with agent-based models with a detailed environment description as important examples.

In this contribution, however, we are occupied with the classification of time series from a long-term ecosystem research site, intended as input to informed decision-making about the most suitable model classes to successfully describe the system behaviour.

## 2. Materials and Methods

*2.1 Site description*

Hydrological catchments or watersheds covered with forests are often used as monitoring units of semi-natural landscapes, not the least since these landscape units conceptually allow for a closed input-output balance for matter fluxes. Such catchments should be small to allow for a homogenous, even-aged forest stand, but they need to be sufficiently large to allow for a perennial stream. If there is only a single perennial stream, the catchment is called first-order or "headwater". The data set of this study is derived from three first-order headwater catchments with an area between 33 and 75 ha within the Bramke valley (Harz, Germany; center coordinates 51.858° N, 10.423° E). The catchments are underlain by fissured Lower Devonian rocks, with deeply weathered soils. They are dominated by Norway spruce stands (Picea abies Karst. L). The observation period covers a full rotation period of the forest stand, i.e. from clearcut to clearcut. For most of this time the catchment was under monitoring of forest growth and hydrochemistry [14]. Since 1994 it also has been a level II site of the ICP-Forests monitoring program [15].

All observations included here stem from three adjacent small headwater catchments in the Harz mountains, known as "Lange Bramke" (LB), "Dicke Bramke" (DB) and "Steile Bramke" (SB) (Fig. 1). At LB, Norway spruce (Picea abies (L.)) is practically the only species (62-70 yrs.), at DB (1.1 % of the area) and SB (14.4 %) replanting with European beech (Fagus sylvatica L.) and European alder (Alnus glutinosa L.) near streams has occurred since 1986. The area has a long history of timber use and charcoal production due to mining, and the Lange Bramke catchment was clear-cut in 1947 as part of reparation payments of Germany to Great Britain after World War II. It was then replanted starting in 1948. Environmental monitoring started the same year, first with discharge measurements (runoff rates) of the stream at LB. After the clear-cut, there were concerns about soil erosion at the steep slopes. The influence of forest cover on the quantity and quality of streamwater was in focus when the Bramke catchment study was implemented in 1948 [16]. Water budgets were of major interest at the time. Starting in the 1960ies, water samples were also taken for water quality assessment. Today, the long-term hydrochemical data allow a nuanced view at internal processes. Here, we used the maximal (biweekly) resolution to characterize four major dissolved ions. In addition, we consider air temperature and runoff coarse-scaled to the same resolution.

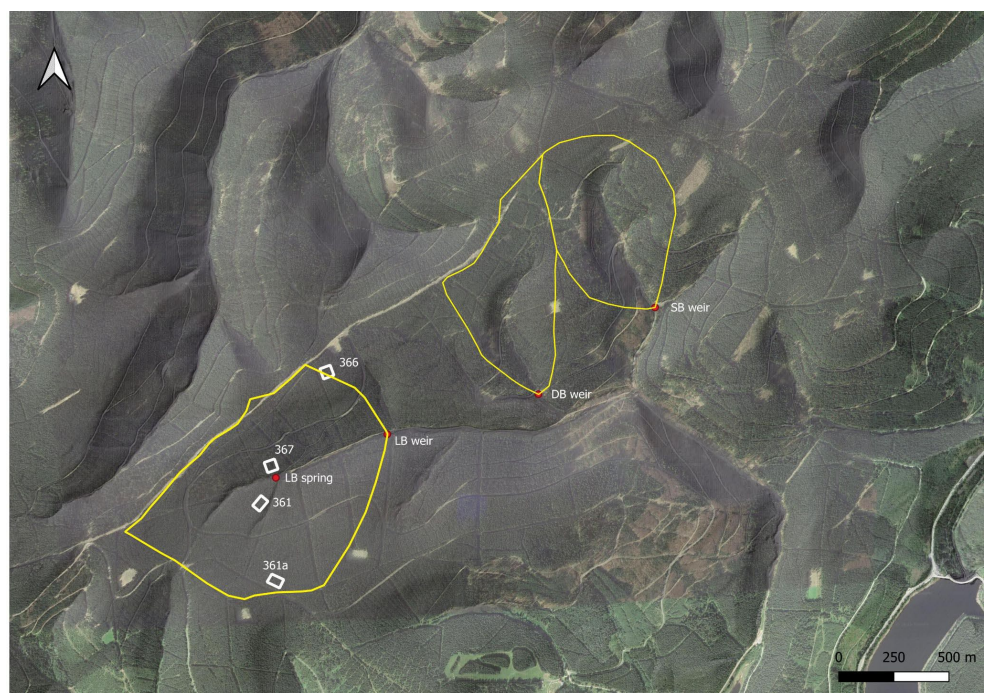

**Figure 1**: The three catchments with four hydrochemical sampling points LB spring (LBQ[1]) & LB weir (LBW), DB & SB weir (DBW & SBW). Rectangles are the forest inventory plots, 361, 361a, 366, and 367 are the numbers of the ICP-Forest Level II monitoring sites. Meteorological data are obtained at the clearing near and northeast of site 366.

The time series analysed here except temperature are all from one of these four sampling points, and we will refer to them as DBW, LBQ, LBW and SBW from now on.

*2.2 Time series from the sites*

---

[1] Lange Bramke Quelle (in German) ≙ Lange Bramke Spring

Water samples from all four sampling points (cf. Fig. 1) are taken at regular two weeks intervals and analyzed for chemical composition. Among the many ions determined in the chemical analysis, we select only four: sulfate ($SO_4^{2-}$), nitrate ($NO_3^-$), chloride ($Cl^-$), and potassium ions ($K^+$). Chloride is part of the atmospheric deposition, reaches the soil dissolved in rain and in throughfall washed off from the needles; it is not actively processed by plants and does not react with the soil matrix. It is thus considered as "ecosystem-inert". This is very contrary to sulfate with its intricate impact on soil chemistry through both bacterial reduction and inorganic sorption processes, while nitrate, as part of the plant nitrogen cycle, is also processed by soil bacteria (process of denitrification), but is not well retained in the soil matrix and can reach the groundwater. Potassium is a major plant nutrient and can be retained in the soil. Thus, these variables are important representatives for plant nutrition or are reflecting soil chemical properties, i.e. they represent different influences from either physical or biological processes on runoff hydrochemistry, or both, and are therefore expected to exhibit different dynamics.

As these chemical concentrations in streamwater are influenced by precipitation, temperature, radiation, and other variables, we expect them to display both long-term trends and an annual cycle (seasonality). The impact of these two deterministic properties of the time series on the complexity metrics will be investigated by comparing the original time series with versions where the trend or the seasonality has been removed.

Water sampling started in the LB catchment at the end of the 1960ies and in the 1980ies for the other two catchments DB and SB. However, due to irregular sampling in the beginning and also some changes in the chemical-analytical methods over the decades, we extracted data from all four sampling points for a common period of 33 years, 1991 to 2023. Due to certain irregularities in the sampling intervals and some gaps contained in the records, we also decided to use the data at regular 14-day intervals, averaging all observations within each given period when there were more than one obtained.

We supplement the ion concentrations by time series for air temperature, taken from a meteorological station within a clearing near the LB catchment (Fig.1) at 2 m height above surface, and with runoff (stream discharge) from the Lange Bramke weir, which is the longest and most continuous record available. Both are obtained at daily resolution but downsampled for our purposes to 14 day resolution.

The original dataset used in the analysis thus is a set of 18 time series (four ions at four samping locations, plus air temperature plus runoff at one location) of length $N = 860$ values each.

Over the lifetime of the forest stand (1948-2022), two major environmental changes occurred at the Harz: Increase and decrease in deposition of air pollutants by long-range transport (acid rain), and climate change. These phenomena are reflected in the trends for some of the variables, most prominently in the decline of sulphate as a result of substantially reduced atmospheric deposition of sulfur dioxide ($SO_2$). We also mention that the catchments were recently severely affected by stand-replacing bark beetle attacks. The planted species was considered as well-adapted to the site conditions, i.e. able to respond and survive the environmental conditions to be encountered over its planned rotation period of 120 years. However, the stand was practically killed (83 % of the forest in the catchment) by a bark beetle epidemic after 71 years only. We expect a dramatic response of stream chemistry due to this damage; this is already seen for nitrate concentrations in the last two years reported here, 2022 and 2023.

*2.3 Data preparation and analysis methods*

Many of the methods applied to our time series here require gap free data, or at least are biased or difficult to interpret when gaps in the data are present. We thus spend some efforts to generate time series at a completely regular temporal resolution with no values missing. We conceptualized each time series as additively composed of trend, periodic, and noise components, where the latter include a potentially complex mix of correlation structure. We will compare the partial series - e.g. the trend component only, or the original series detrended - and thus need an operational method to decompose the series into these three components. We use a general notion of a trend as any "static" component, i.e. having no identified periods and no high-frequency noise present. Given the temporal resolution, and the time span covered, the only periodic component notoriously present is the annual cycle. Among the existing methods for decomposition, we choose a fully data-adaptive flexible method which at the same time can be utilized for gap filling as well: the Singular System Analysis [17].

*2.3.1. Gap Filling, Detrending and Deseasonalization: Singular System Analysis*

Annual cycles are notorious for most of the water chemistry variables, basically induced by yearly cycles of temperature and radiation, but also crucially determined by the biological activity during the growing season and hibernation during winter. The presence of trends is due to a mixture of the growth of the forest stands - the observation period is a significant fraction of the average lifespan of a spruce stand after plantation - and nonstationary environmental conditions (atmospheric deposition of air pollutants, climate change, disruptive events). As a result, our typical time series would not pass any classical stationary tests.

Some of the 14 day intervals did not contain a single value for some of the ions, i.e. the original time series also contained some gaps. As most of the methods require gap-free data to avoid bias, one ought to fill the gaps prior to their application.

The following figures show gap-filled versions of the time series, representing the whole data set going into all further analysis.

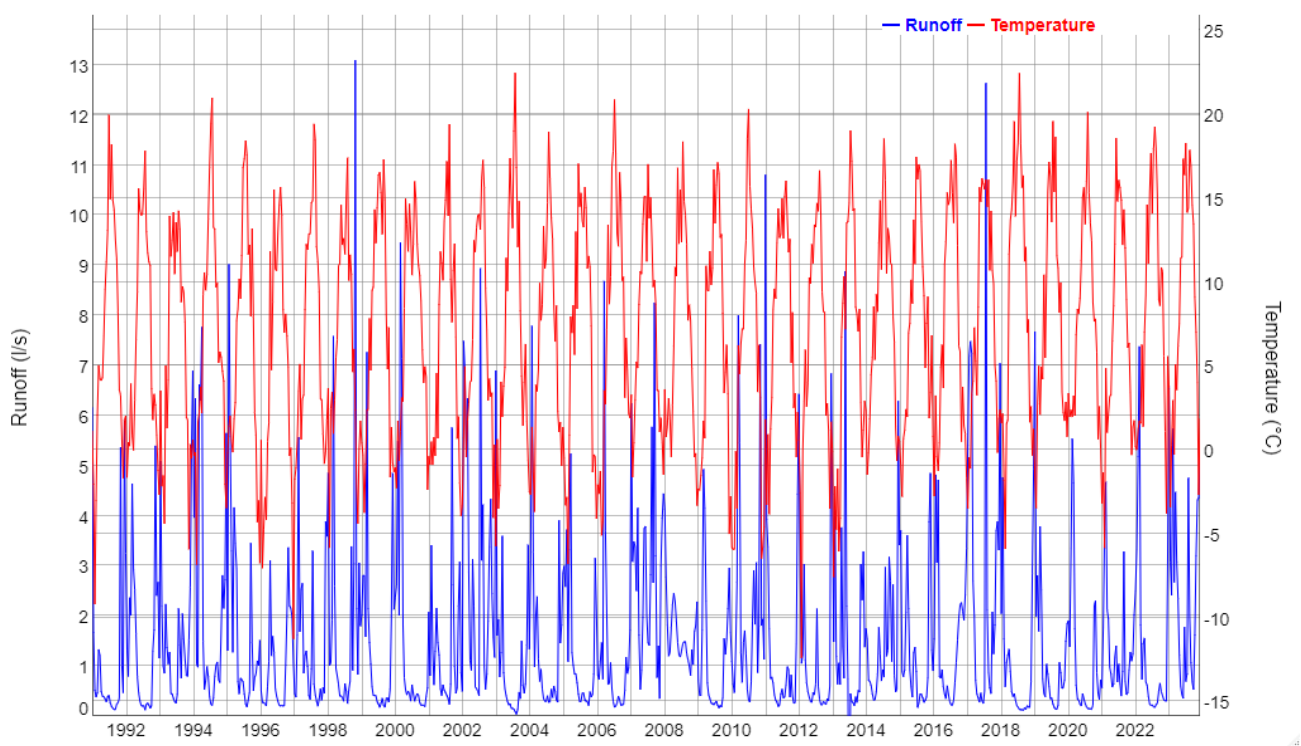

**Figure 2**: Runoff at the weir of the Lange Bramke catchment (LBW) (blue, left axis) and air temperature at the meteorological station within the catchment (red, right axis), values aggregated to 14 days based on daily observations, 1991 - 2023.

Fig. 2 shows runoff at the gauged weir at Lange Bramke. The LB stream is perennial, but almost ceases in the very dry summers of 2003 and 2018. The temperature record (14-day minimum: -12.4 °C, 14-day maximum: 22.4 °C) has a highly significant positive trend in the observation period with a slope of +0.062 °C / year, which is however difficult to recognize in the Figure.

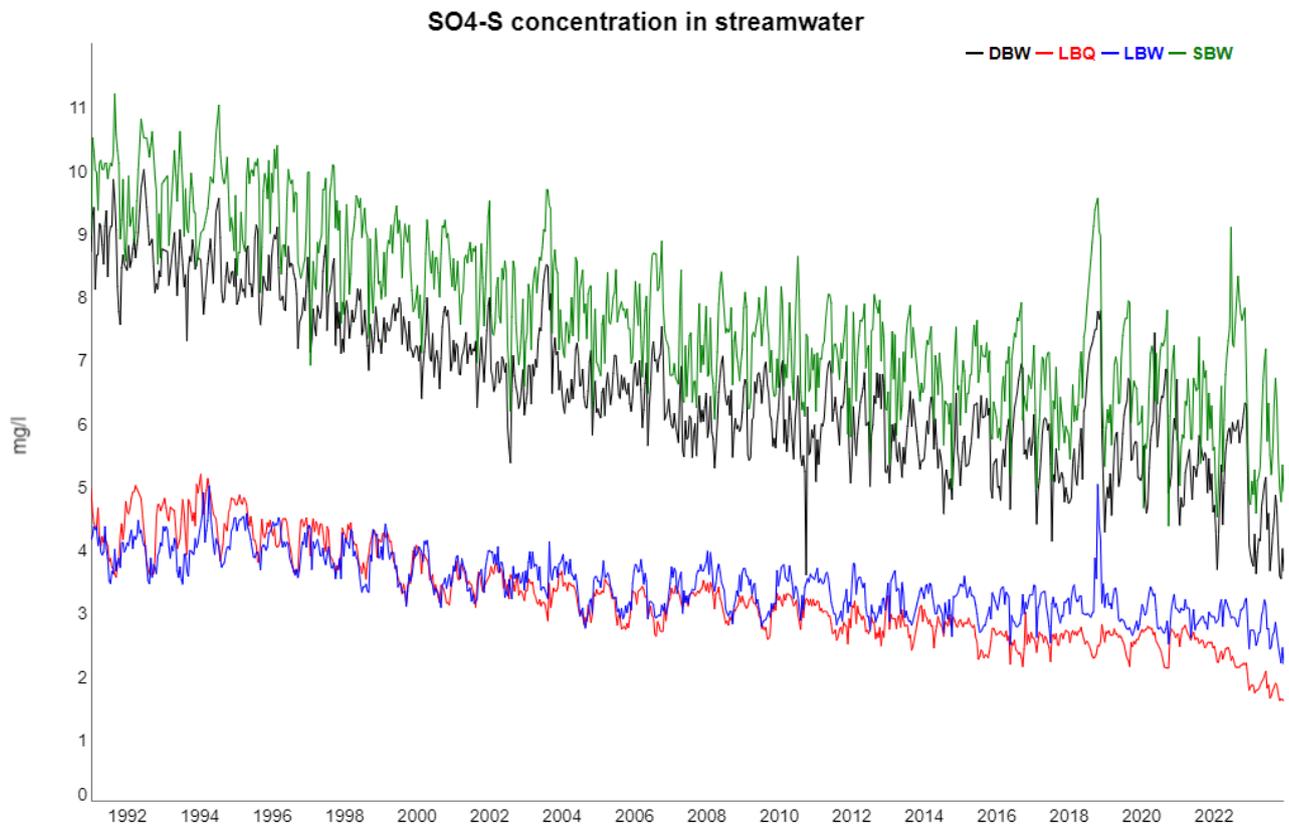

**Figure 3**: Sulfate concentrations at the four locations DBW, LBQ, LBW and SBW. The clear decreasing trend is due to recovery from acid deposition since the late 1980ies.

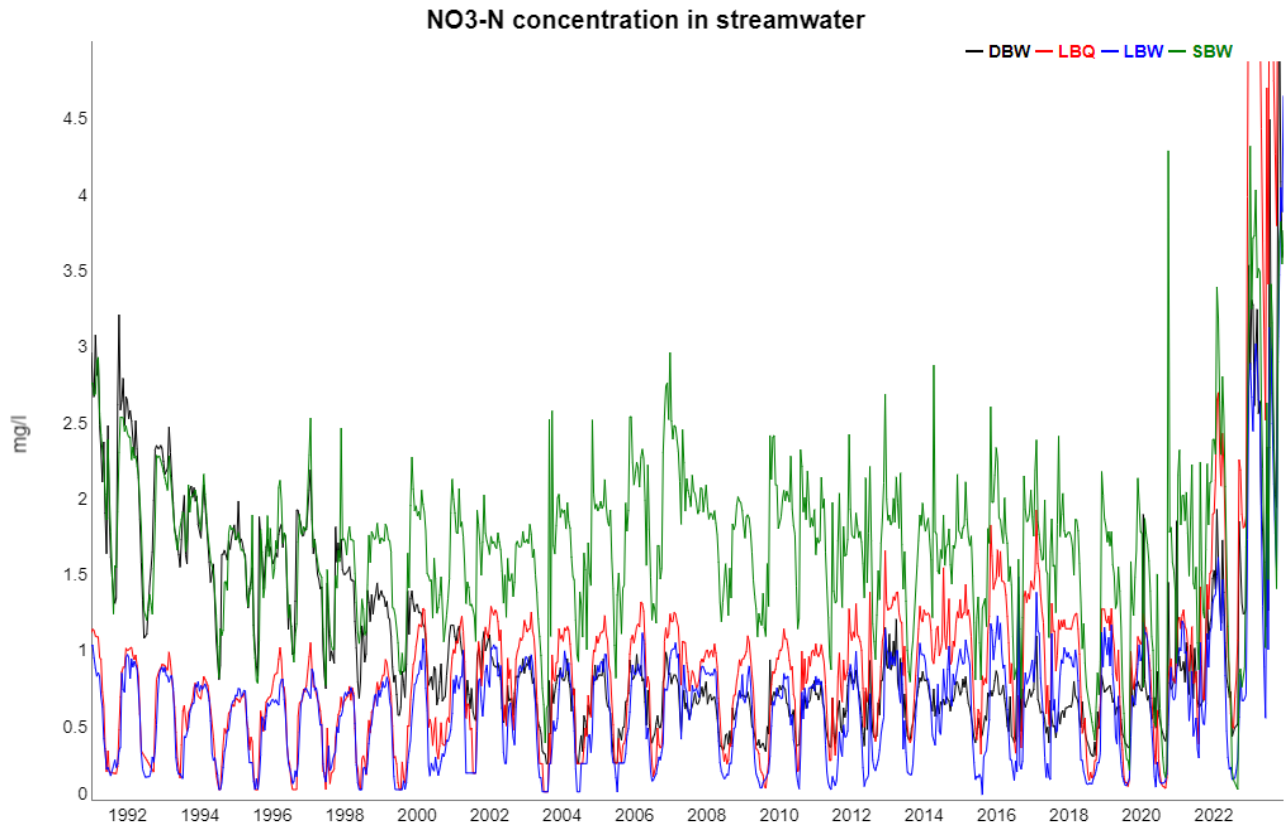

**Figure 4**: Nitrate concentrations in streamwater at the four locations during 1991 - 2023. The sharp increase since 2022, exceeding even the concentration range displayed here, is due to forest dieback induced by a bark beetle attack.

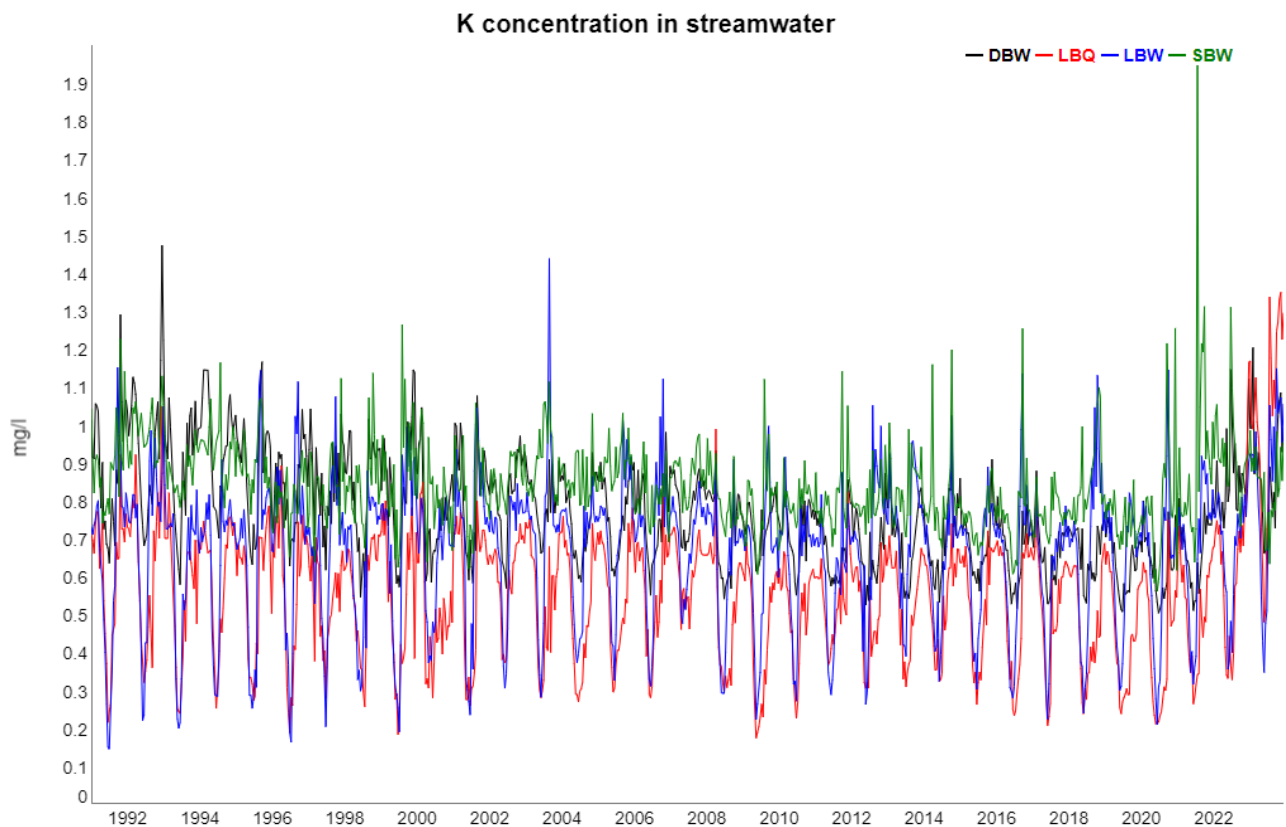

**Figure 5**: Potassium concentrations in streamwater at the four locations 1991 - 2023.

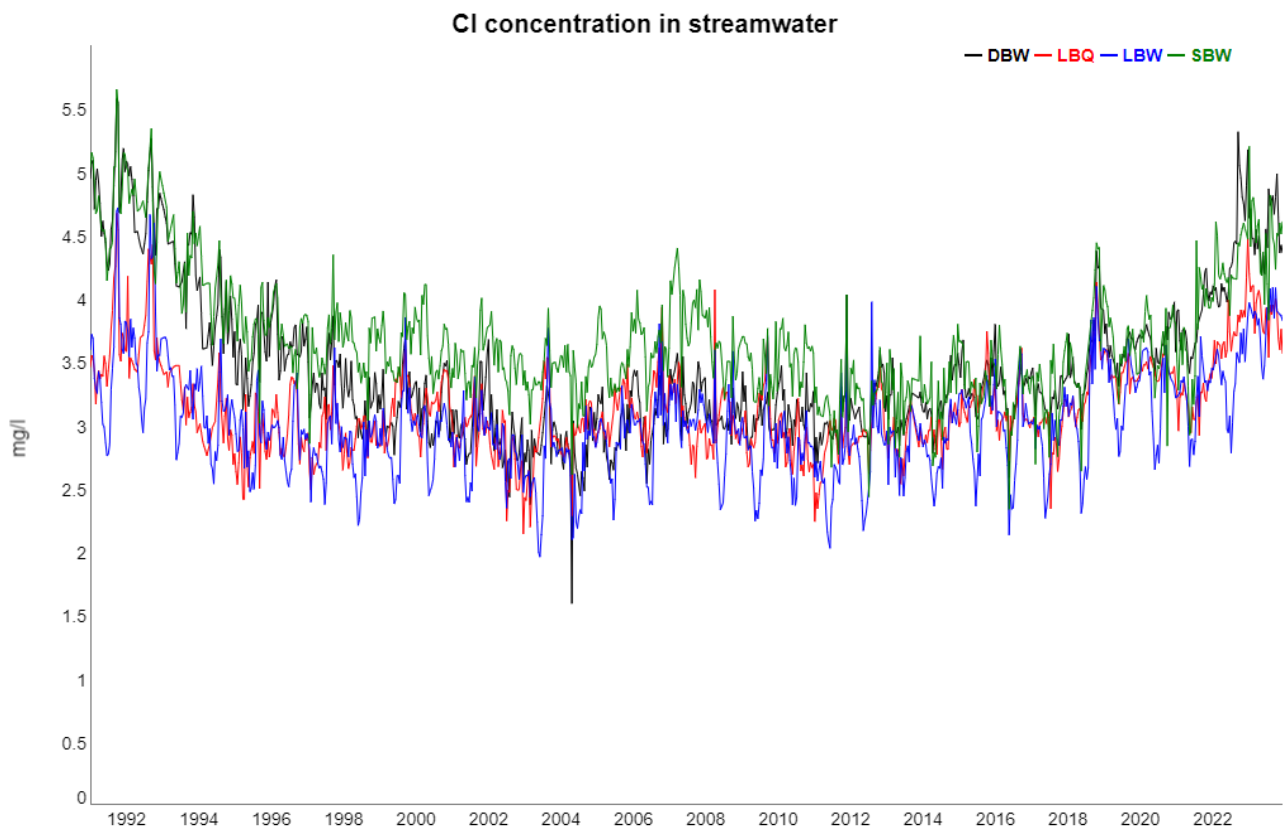

**Figure 6**: Similar to Figures 2 to 5, here for Cl concentrations.

Our tool to deal with all three issues of gap filling, detrending and deseasonalization is the Singular System Analysis (SSA) in the version of [18], i.e. a fully data-adaptive decomposition method based on the lagged covariance matrix. SSA leads to an orthogonal set of eigenmodes, ranked according to their explained variance, which are the eigenvalues of a Singular Value Decomposition. SSA comes with one parameter, the embedding dimension or window length $L$. No strict rules exist to find an optimal value of $L$; however, $N/L \approx 2 - 10$ is recommended. Periodic components appear as pairs of eigenmodes with nearly identical eigenvalues, and single components with no detectable period smaller than $L$ (quasi-static) are considered as a trend. The periodic components are not necessarily sinusoidal, nor is the trend necessarily monotonous.

We used the R package R-SSA [19] to decompose the time series. The SSA also allows filling in missing data with several methods; in our case, we are using the "Caterpillar" algorithm [20] which requires selecting a group of SSA components to base the gap-filling on. This group should contain trends and major periodic components as a minimum. In our case, we selected the components with the six highest ranks, the ones with the annual cycle and the trend always among them. Isolated single missing values are eliminated by simple linear interpolation.

After gap-filling, the complete time series are used to identify the components containing the trend and the ones with the annual cycle. An overview of their contribution to the total variance of the time series is provided by Tab. 1. This leaves us with six versions of the respective time series: the original one, a detrended one where the trend component is subtracted but the annual cycle is still contained, one where the annual cycle is removed but the trend is still there, the trend and the annual cycle alone, and the residual time series where both the trend and the annual cycle has been removed. Further analysis is conducted on all six versions and then compared.

Table 1: Percentage of explained variance relative to the total variance of the original time series for the variables investigated, based on SSA decomposition.

| **Variable** | Process | %Variance per Location | | | |
|---|---|---|---|---|---|
| Temperature | Trend | 0.97 | | | |
| | Season | 80.99 | | | |
| | | DBW | LBQ | LBW | SBW |
| Runoff | Trend | - | - | 0.48 | - |
| | Season | - | - | 20.69 | - |
| Cl | Trend | 86.05 | 57.43 | 24.00 | 73.93 |
| | Season | 4.42 | 5.57 | 19.15 | 4.90 |
| K | Trend | 39.74 | 24.51 | 0.76 | 19.41 |
| | Season | 31.84 | 46.11 | 45.89 | 4.48 |
| $NO_3$ | Trend | 72.06 | 74.16 | 45.41 | 20.59 |
| | Season | 7.39 | 11.45 | 30.27 | 32.56 |
| $SO_4$ | Trend | 75.91 | 88.02 | 64.65 | 70.12 |
| | Season | 3.17 | 4.46 | 15.33 | 6.68 |

It is obvious from Table 2 that in most cases, trend and seasonal component combined are containing a large fraction of the total variance. On the other hand, there are major differences between the variables, as is also recognizable from the time series plots: e.g., for runoff and K at LBW, the trend is negligible; for $SO_4$, the trend dominates by far over the seasonal component. $NO_3$ has a very strong trend, but that consists of a decline during the 1990ies and an increase in the last years, so it is non-monotonous.

The percentages reported in Tab. 1 do not reveal whether this periodic component is synchronized between the variables, or if they show different phases but with a constant phase relation. Insight into these connections can be gained by extracting the seasonal component only (usually a pair of eigenmodes of the SSA decomposition) and then calculating the instantaneous phase using the Hilbert transformation of the time series. Time series of the difference between the two instantaneous phases are constructed, and interpreted as lag between the two annual cycles by converting it to time scales. The degree of synchronization, a measure for the stability of the phase relation between the two, can be obtained by calculating the Mean Resultant Length [21].

The SSA decomposition is illustrated in Fig. 7, using Cl at LBW as an example. For most variables, removing the seasonal component, or the trend, or both to obtain the residual, does not change the visual appearance of the time series substantially. This observation indicates that the stochastic component is dominating the dynamics. The extracted trend and seasonal components separately (lowermost two time series in Fig. 7) exhibit rather smooth and regular dynamics, and are the deterministic part of the time series in this framework of additive SSA decomposition.

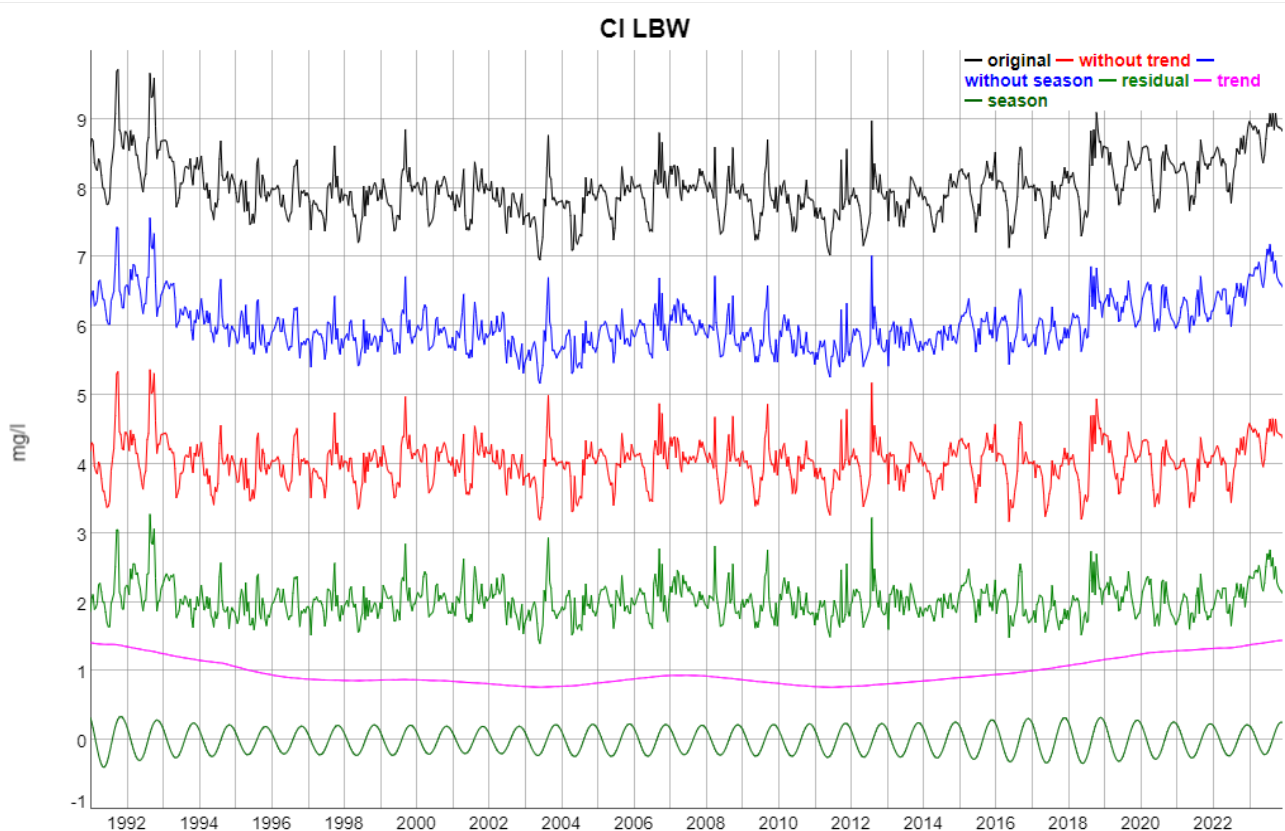

**Figure 7**: Example of the SSA-based time series decomposition: Cl concentrations at LBW. The individual time series have been shifted for easier identification; e.g. the original time series was increased by a constant of 5 mg/l. From top to bottom: original time series; original minus annual component; original without trend component; residual = original - trend component - annual component; the trend component alone; and the annual component alone.

Note that the trend component generated by SSA is a nonlinear, non monotonic function.

*2.3.2. Permutation Entropy and Complexity*

We use symbolic dynamics for quantifying entropy and complexity of the time series. Following the seminal approach of [22], we construct ordinal patterns from the real-valued series. The crucial parameter to do that is the embedding dimension (aka word length or pattern length) $D$. Considering that in order to capture as much structure on different timescales as possible, one wants to maximize $D$, but the factorial explosion of the number of possible patterns quickly prevents statistical saturation for a given time series length $N$, we respect the rule of thumb that $N > 5D!$ should hold and thus fix $D = 5$. This implies that the temporal distance between the first and the last value in any pattern is roughly two months (precisely 56 days). The regularity of the annual cycle is undetectable with these settings. Ordinal pattern analysis is strongly dependent on temporal resolutions.

From the ordinal pattern probability distributions (opd's), we calculate the Shannon entropy, in this context also known as the permutation entropy, and the permutation complexity, also known as MPR complexity, following the seminal work of Rosso and co-workers [8,23]. Permutation complexity is based on the Permutation Jensen-Shannon divergence [24,25]

$$JSD(p,q) = S(\frac{p+q}{2}) - \frac{1}{2}(S(p) + S(q)) \tag{1}$$

with $S(p)$ the opd-Shannon entropy and with white noise with its equidistributed opd as reference process ($q$). The square root of $JSD$ can be shown to be a proper metric [25].

The entropy-complexity plane contains regions which are unreachable for any time series; the accessible area is delimited by a lower and an upper limit curve, with their shape depending on $D$. It also turns out that power law noise, i.e. correlated noise whose power spectral density scales as $P(f) \sim f^{-k}$, forms a single one-dimensional curve in this plane. This curve will be a reference to judge on the type of stochastic process we observe in our time series.

### 2.3.3. Fisher Information

As third metric quantifying an information-related property of time series, we consider the Fisher Information adapted to ordinal pattern series [12]:

$$F(P) = \frac{1}{2}\sum_{i=1}^{D!-1} [\sqrt{p_{i+1}} - \sqrt{p_i}]^2 \tag{2}$$

where $P$ denotes an ordinal pattern distribution and the $p_i$ are the probabilities of the patterns. The Fisher Information is not a unique quantity since a numbering scheme for the patterns is required, inducing an ambiguity which is, however, insignificant for interpretation purposes as our experience indicates. In this work, we use the coding scheme of Karsten Keller [6], and $D = 5$ as before. A two-dimensional plot of Fisher Information versus Permutation Entropy is not known to have limit curves, i.e. every point in the square [0,1]x[0,1] is reachable, and positions of time series in this plot can be used to conclude on stochasticity, and to compare to reference processes.

### 2.3.4. Renyi and Tsallis entropy and complexity

Permutation entropy and complexity can be considered as special cases of a class of entropies and complexities, introduced by Rényi [26] and Tsallis [27], respectively. Either class is parametrized through a non-negative real number; within the concept of Renyi, the parameter $\alpha$ is used to apply power weights to the probabilities of eq. (1), either enhancing ($\alpha < 1$) or suppressing ($\alpha > 1$) rare patterns. The Renyi approach puts power weights to the probabilities of the ordinal pattern distribution for the calculation of the entropy:

$$H_\alpha(p) = \frac{1}{(1-\alpha)\ln D!} \ln \sum_{i=1}^{D!} p_i^\alpha \tag{3}$$

In the case of the Tsallis entropy, its parameter is usually called $q$, a generalized logarithm $log_q$ is introduced, and the Tsallis entropy is defined as

$$H_q(p) = \frac{1}{D!}\sum_{i=1}^{D!} p_i \, log_q \frac{1}{p_i} \tag{4}$$

Both entropies converge to the usual Shannon / Boltzmann one if their parameter is chosen to be one. Only the Rényi entropy shares the property of extensivity, i.e. additivity in the case of independent distributions, with the conventional entropy.

Starting from these entropy generalizations, corresponding complexities were defined, formed as the product of the respective entropies with appropriate distance measures to a reference process (white noise), normalized with a maximum distance.

The resulting Rényi complexity[28] allows for a qualitatively distinction between stochastic and deterministic-chaotic time series; for the former, in a entropy-complexity plane, varying $\alpha$ leads to a monotonous behaviour of both entropy and complexity which is not the case for well-studied deterministic maps. In the case of the Tsallis complexity [29], while the deterministic processes lead to open curves with two ends, the stochastic ones form closed loops when running through all q values. The area covered by the loops is related to the Hurst parameter for the stochastic series.

*2.3.5. Tarnopolski diagrams*

Another method to locate our time series in the context of standard reference processes is provided by the Tarnopolski diagram [30]. Here, one plots two rather simple and parameter-free quantities of each time series against each other: the number of turning points $T$ and the sum of squared differences of adjacent values, also known as the Abbe value $\mathcal{A}$. Reference stochastic processes like fBm or fGn build invertible functions in the $T$ - $\mathcal{A}$ plane Properly normalized, they are largely insensitive to the exact time series length; but still, Tarnopolski found exact equations for the two reference processes fBm and fGn depending on the Hurst parameter at any time series length [31].

*2.3.6. Horizontal Visibility Graphs*

A conceptual simple geometric visualization of the correlations (in a rather general sense, not restricted to linear ones) is the question: sitting on a point in the time series, how far could you see in horizontal direction before other (higher) values are blocking your view? This is the idea behind Horizontal Visibility Graphs (HVGs) [32], a member of the family of complex networks. The resulting network of visibility is analyzed e.g. through its degree distribution; for some processes, it is known that the probability of finding a network node with degree $k$ is exponentially decaying:

$$P(k) = \frac{3}{4} e^{-\lambda_{HVG} k} \tag{5}$$

which is a robust result independent of the time series distribution [33] ; in the absence of autocorrelations, there is even the analytical result $\lambda_{HVG} = \ln(3/2)$. From the observed degree distributions for our time series, we estimated the slope of the relation (5), i.e. $\lambda_{HVG}$, and compare it in particular to the uncorrelated case, and among the time series.

*2.3.7. Complexity of Ordinal Pattern Positioned Slopes (COPPS)*

Another recent approach to quantify the complexity of time series is combining ordinal pattern statistics and network construction. It is based on the slopes present in the ordinal patterns [34]. Encoding the ranks within a pattern of embedding dimension $D$ as integer numbers, one determines the maximum slope (differences of ranks) within the pattern and its position (ordinal pattern positioned slopes, OPPS). Grouping OPPS together with a grouping parameter $s$ leads to a transition network. The ability of a system to generate new patterns whenever the network depth $s$ is increased is the notion of complexity in this context. Thus, the COPPS variable $\lambda_s(D)$ introduced in length. We compare the COPPS values obtained for our time series with a few of the reference processes.

## 3. Results

For each of the 18 time series, we have six different versions due to the SSA decomposition: (1) the original as obtained from field samples; (2) the one where the trend component has been removed; (3) a set where the respective annual cycles have been removed, but the trends remain; (4) the version where both the annual cycle and the trend have been removed, which we call the "residual" time series; (5) the pure annual cycles, and (6) the pure trend components. Note that the last one is nonlinear and might be complicated, although not complex with our notion of complexity.

We compare the different versions of the time series for each method, focusing on the difference in dynamical structure obtained. We start, however, with a linear method to visualize the correlation between the data sets as a correlogram in Fig. 8.

*3.1. Correlograms, phase shifts and Jensen-Shannon Divergence of the time series set*

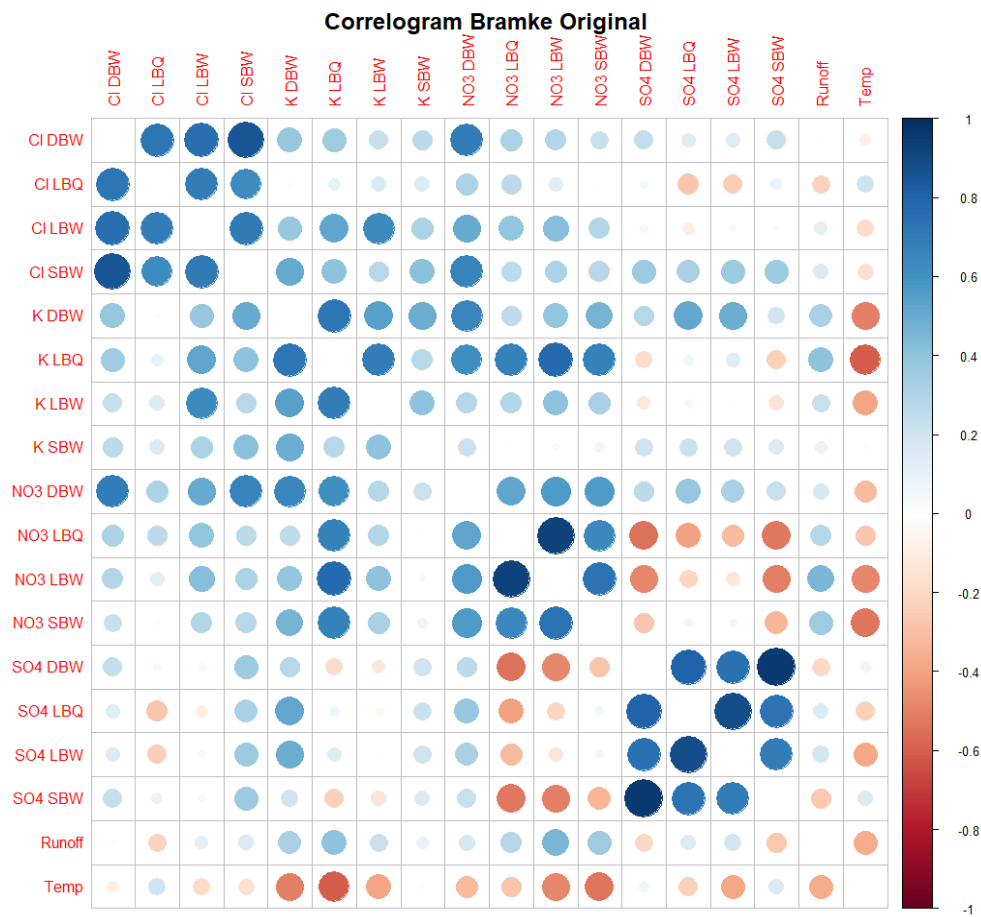

**Figure 8:** Correlogram of the original time series. Both the color as well as the size of the circles indicate the Pearson correlation coefficient of each pair of time series (its absolute

value in the case of the size). The main diagonal (with $R^2 = 1$) is blanked out for better visualization.

The correlogram identifies the group of Cl ions as rather closely connected; much less so for K, apart from the pair of K LBQ / K LBW which are from the same stream. $NO_3$ is again stronger connected, and $SO_4$ even stronger. Since potassium ($K^+$) is strongly processed by plants, this might point to differences in the vegetation dominating the three catchments.

There are also some anti-correlations, e.g. between $NO_3$ and $SO_4$, but also between runoff and K, $NO_3$ and $SO_4$. Given the strong seasonality, this indicates that these ions are not in synchrony, there is a lag of several months between the different pairs.

Correlograms for other variants (trend only, annual cycle only, residuals) are shown in the Appendix (Figs. A.1 to A.3).

This expectation is partially confirmed through the phase shift analysis, where we calculate the instantaneous phase between the annual components of two variables, recalculated to a time lag bound between -6 months and +6 months.

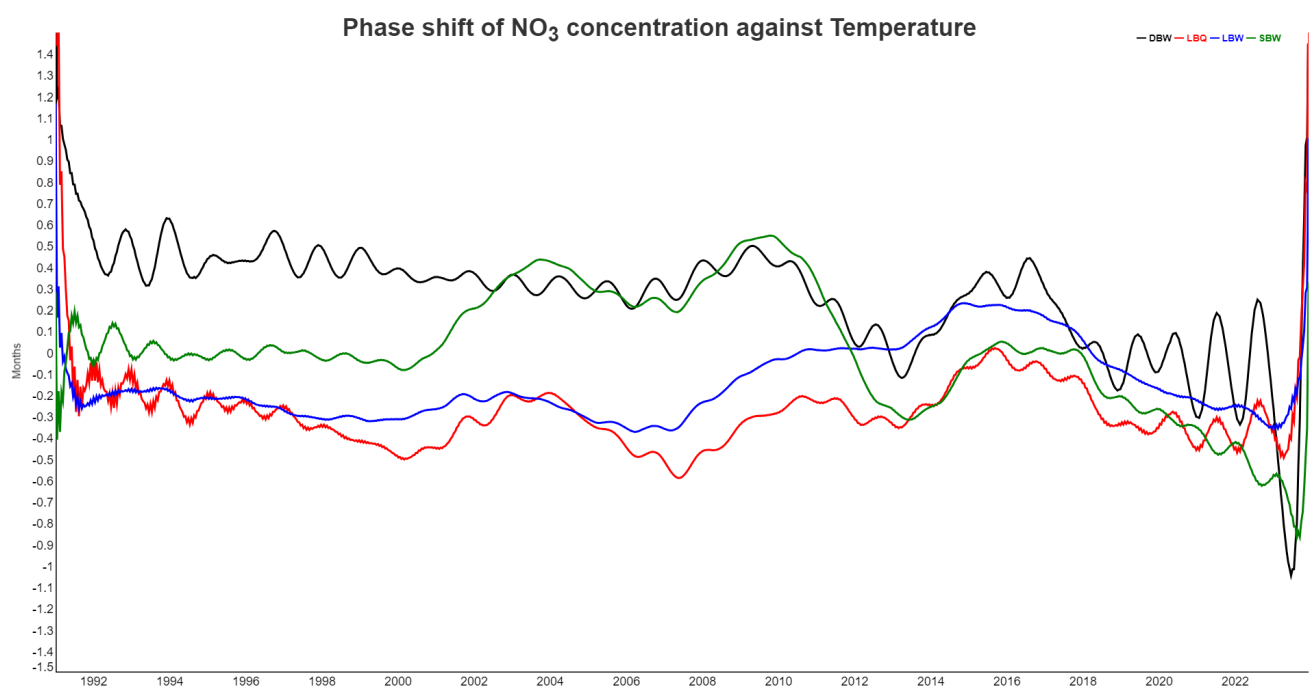

**Figure 9**: Phase shift between $NO_3$ concentrations and air temperature. For better visualization, temperature went into the instantaneous phase calculation with reversed sign, which basically means a shift by six months.

In Fig. 9, the time lags between $NO_3$ concentration from all four locations and temperature are shown. The proper time lag revolves around 6 months, and flips around between +6 months and -6 months which means the same shift for annual cycles. Nitrate in runoff is high when root uptake and temperature are low and we used the negative temperature instead. The proper interpretation of Fig. 9 is therefore that nitrate and temperature are half a year apart, easily explaining the negative correlation coefficient seen in the correlogram. The phase shifts are reasonably stable most of the time, although they also contain periodic components of unknown origin. In the last two years, $NO_3$ and temperature decouple from each other due to the tree mortality, so the biological control on nitrate largely disappears.

Fig. 10 shows the phase shift between $SO_4$ at SBW and $NO_3$ at all locations. There is no obvious coupled cycle between the two, but both ions are dominated by biological processes and also through soil interactions. For LBQ, LBW and SBW, the sulfate signal comes mostly first, but the lag is rather small, typically less than a month. For DBW, however, $NO_3$ is completely out of phase with lags around 6 months, the biggest possible. Again, in the last two years, $NO_3$ is decoupling from the $SO_4$ dynamics.

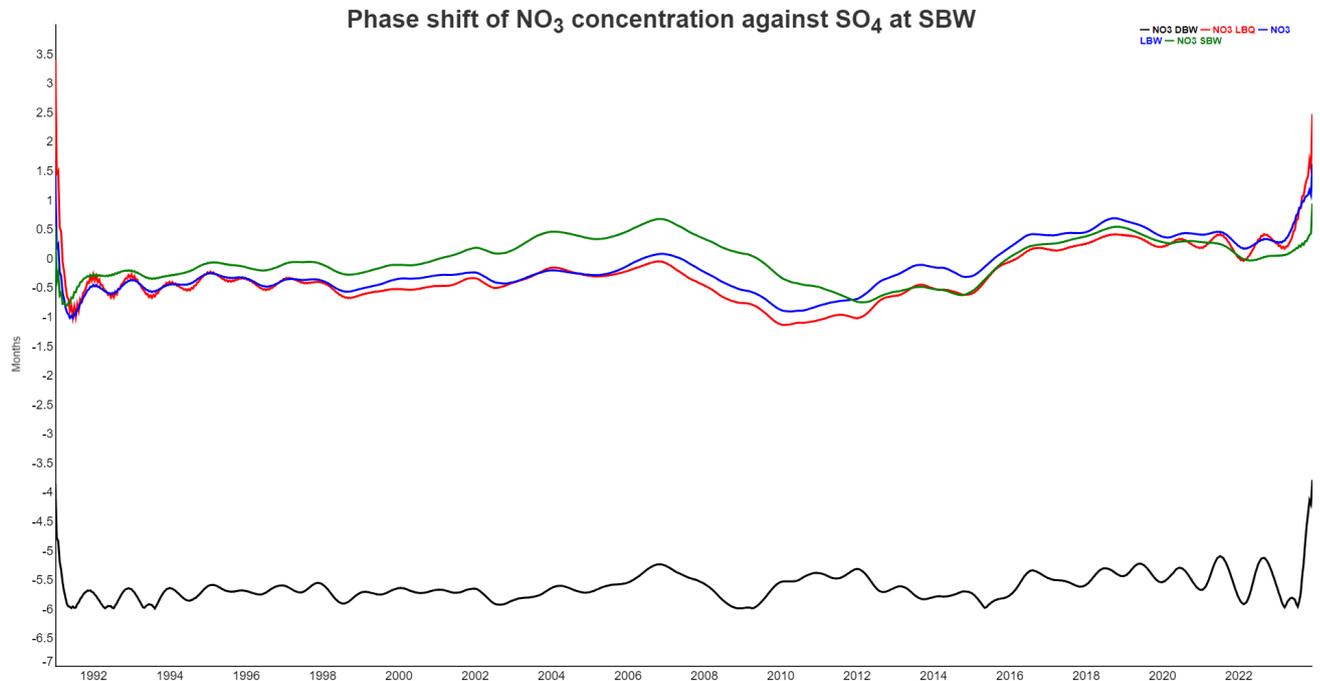

**Figure 10**: Phase shift between $NO_3$ at all locations and $SO_4$ at DBW. Positive values imply that $SO_4$ comes first in the annual cycle. Since the phase shift is cyclic (minus six months is equivalent to plus six months), we used the negative absolute value for the phase in the case of $NO_3$ at DBW, which would otherwise flip around frequently.

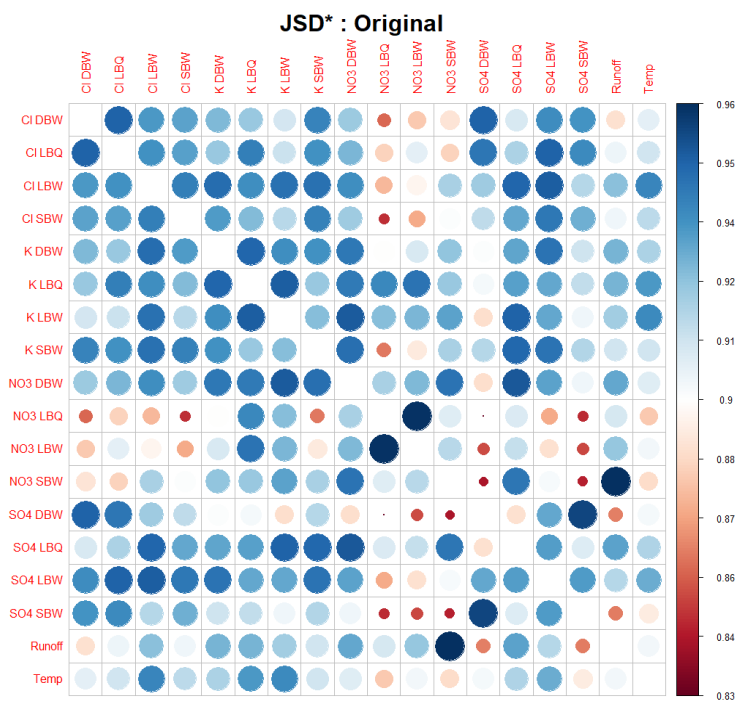

**Figure 11**: Complementary Jensen Shannon Divergence for the original time series. The trivial main diagonal entries are suppressed.

As an alternative approach to visualize the connection between the different variables, we calculate the Jensen-Shannon Divergence of the respective ordinal distributions. As this is a distance measure[2] with $JSD(p,p) = 0$, and we want to compare to the correlogram, we use the complementary $JSD^*(p,q) = 1 - JSD(p,q)$ with $0 \leq JSD^*(p,q) \leq 1$, and values close to 1 if the two distributions are rather similar, as with the correlation coefficient.

Fig. 11 displays $JSD^*$ for the 18 time series. Note that we are comparing pairs of temporal structures, as expressed through the ordinal pattern distributions at $D = 5$. Synchronicity of pattern occurrences is not part of the comparison directly; you might shift one time series relative to the other and still get the same result. There are three pairs of variables which are very close to each other in the ordinal pattern space: $NO_3$ at the two locations LBQ and LBW (which is the same stream); $NO_3$ SBW and Runoff; and $SO_4$ at SBW and at DBW. Some of the pairs which are Pearson-anticorrelated appear with small values for $JSD^*$ in Fig. 11. The pair ($SO_4$ LBW, $NO_3$ DBW) which showed this exceptionally behaviour in Fig. 10 has a low but not exceptional value for $JSD^*$.

### 3.2. Entropy-Complexity plane

Permutation Entropy (PE), Permutation Complexity (MPR) and Fisher Information (Fis) were calculated with $D = 5$. For the PE versus MPR plane, results are shown in Fig. 12.

---

[2] We are aware that a proper metric is only obtained when using the square root of JSD instead. This is however not a relevant aspect for our application here.

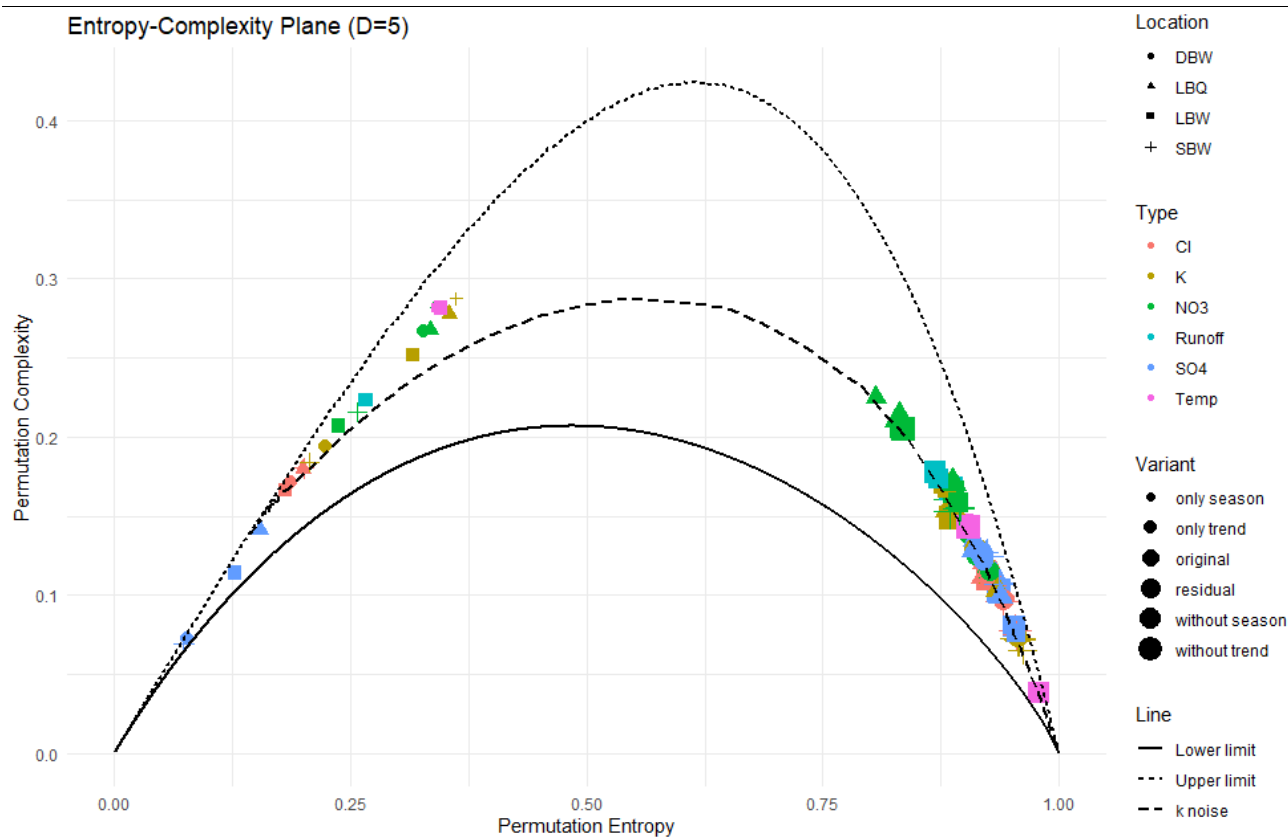

**Figure 12**: The entropy-complexity plane (PE, MPR) for our time series. The plot symbol indicates the location, the size of the symbol corresponds to the variant ("original" to "trend only"), and the color refers to the variable observed. The upper and lower limit curves for $D = 5$ as well as the powernoise curve are also shown.

Most of the time series are close to, or even on, the k noise curve. The removal of the trend and/or the annual component let them move a bit upwards to lower entropy and higher complexity, but still along the powernoise curve. The trend and annual component alone, however, occupy an area on the left side of the maxima of the curves, with corresponding low values for the entropy. In particular the "trend only" variant exhibits higher complexity values than the powernoise would indicate, demonstrating the less stochastic nature of the trend component. The opd´s of these time series also have a lot of missing patterns.

## 3.2. Entropy- Fisher Information plane

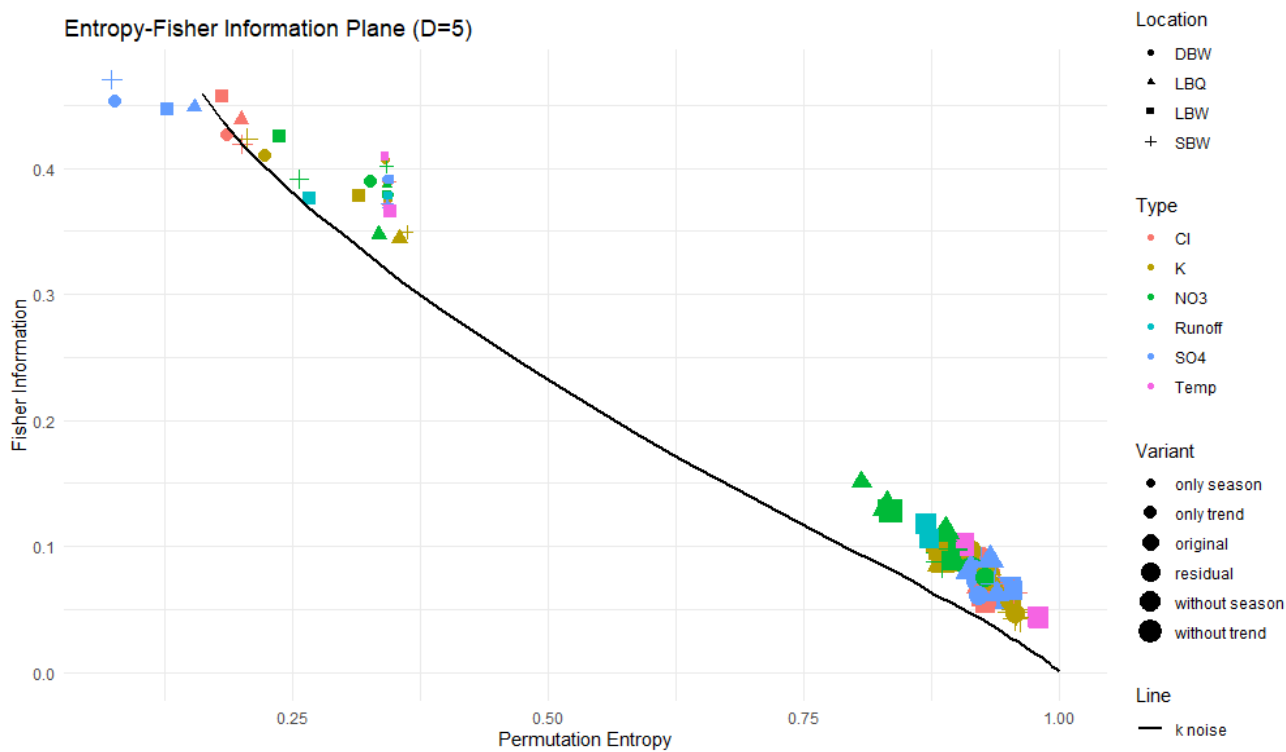

**Figure 13**: The Fisher Information-Entropy Plane for our time series. For the legend, see Fig. 12. The curve for k noise is plotted for reference.

The conclusion from Fig. 12 that the original and detrended series are quite compatible with k noise cannot be drawn in the same manner for the Entropy-Fisher Information plane. Again, the trend and annual component alone are very distinct from the other variants, but for the latter, every time series shows higher Fisher Information than the k noise. This indicates that the opd's for them are more heterogeneous; in particular, some of the patterns might occur very rarely or not at all at that time series length, which is contrary to the k noise where all patterns appear. In fact, each of our time series exhibits missing patterns in all variants, resembling deterministic parts still contained in them.

### 3.4. Rényi and Tsallis entropy-complexity planes

Varying the Rényi parameter $\alpha$ in a wide range (we chose $0.01 \leq \alpha \leq 100$) ensures that the whole range of possible enhancements or discriminations for the pattern probabilities is covered. The result is one curve per variable per treatment. This is shown in Figs. 14 and 15. For the original time series (Fig. 14) and the three variants where part of the signal is removed (not shown), all curves have the common property that Rényi complexity is decreasing with increasing Rényi entropy. This is a behaviour characteristic for correlated stochastic processes. The opposite is true for the trend and the annual component (Fig. 15), as these are resembling deterministic processes.

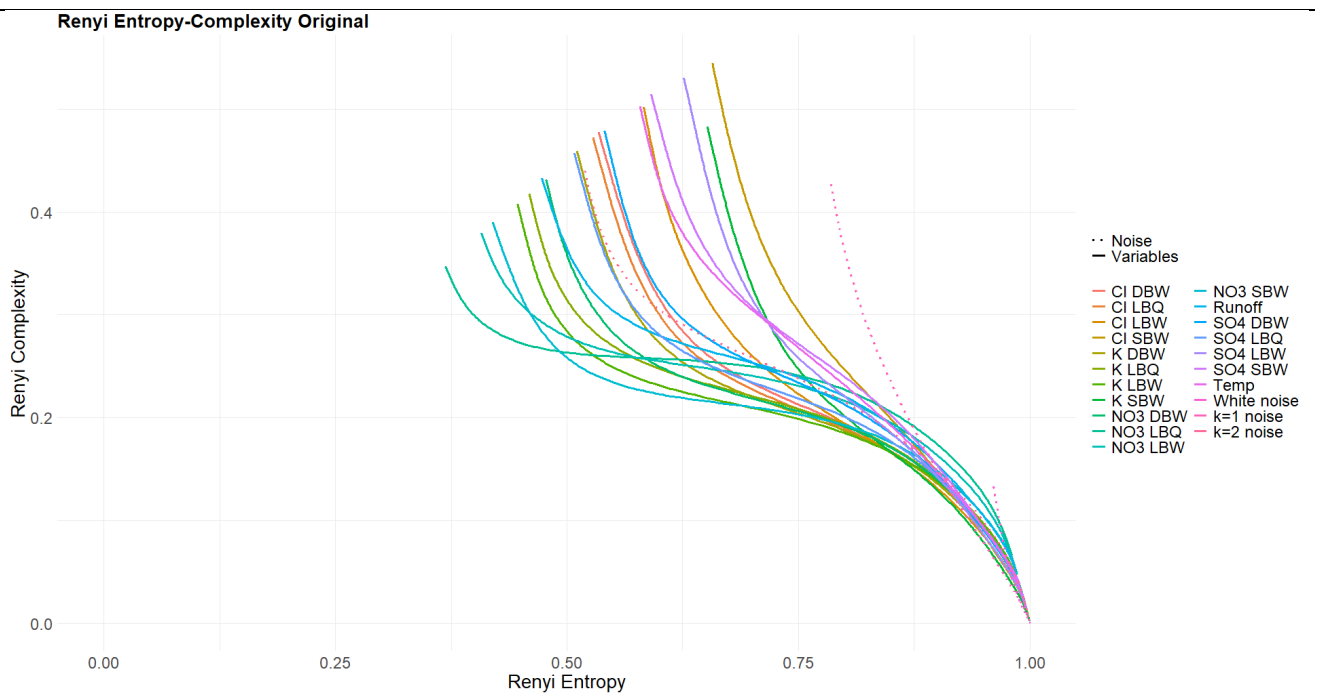

**Figure 14:** Renyi entropy-complexity plane for the original time series. Three powernoise processes are displayed for comparison.

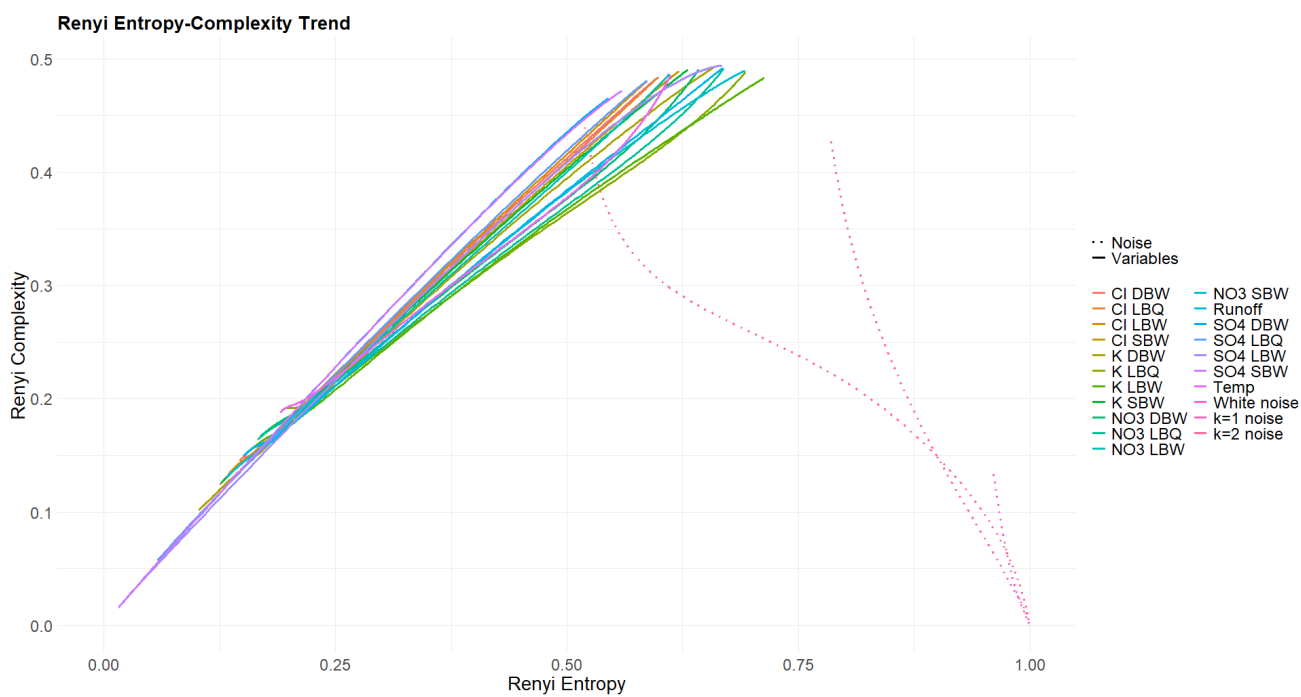

**Figure 15**: Renyi entropy-complexity plane for the trend components. Three powernoise processes are displayed for comparison.

The Tsallis parameter $q$ also was selected from the interval $0.01 \leq q \leq 100$; note that $q = 0$ is a pathological case. Figs. 16 and 17 show the $(H_q, C_q)$ planes for the original time series and for the annual components, respectively. Note that none of these curves are closed loops, contrary to the powernoise reference processes (the white noise (k=0) curve is actually just a point at $(H_q, C_q) = (1,0)$ independent of $q$). This is due to the presence of missing patterns: [29] have shown that the Tsallis curves start in our case at $(H_{0^+}, C_{0^+}) = (\frac{119-m}{119}, \frac{m(119-m)}{119^2})$ when $q \to 0^+$, where $m$ is the number of missing patterns.

The latter are notorious for deterministic time series, but also occur for stochastic processes, depending on the embedding dimension [6]. Still, the original time series curves for $H_q$ and $C_q$ resemble very much the loops of stochastic processes (Fig. 16), whereas the annual component does not (Fig. 17). They are clearly open curves. For their endpoints for $q \to \infty$, [29] provide an analytical equation, which can be checked against the value obtained for the largest $q$ used; our choice $q = 100$ is clearly sufficiently asymptotic.

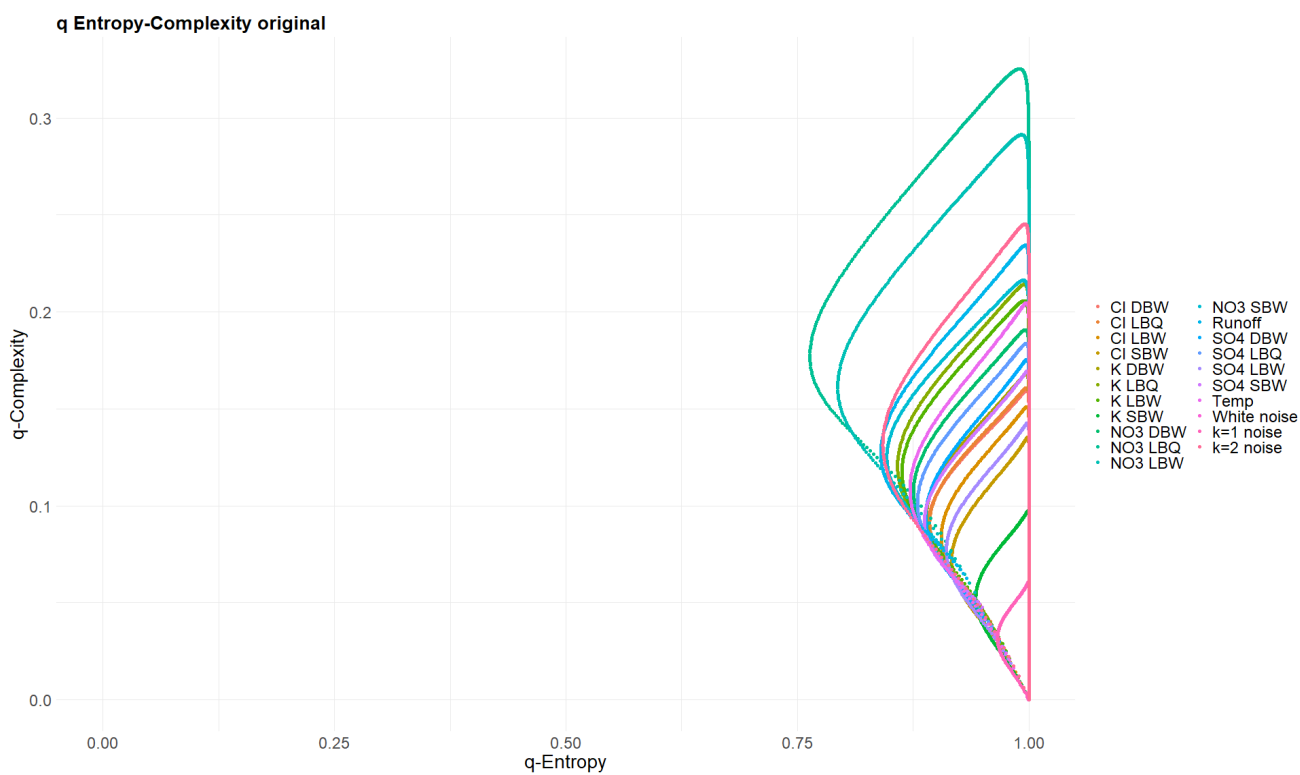

**Figure 16**: Tsallis q-Entropy versus q-complexity for the original time series. The scaling of the entropy axis is deliberately chosen to cover the whole possible interval [0,1]. Powernoise curves for k = 0, 1 and 2 are indistinguishably part of this set of curves.

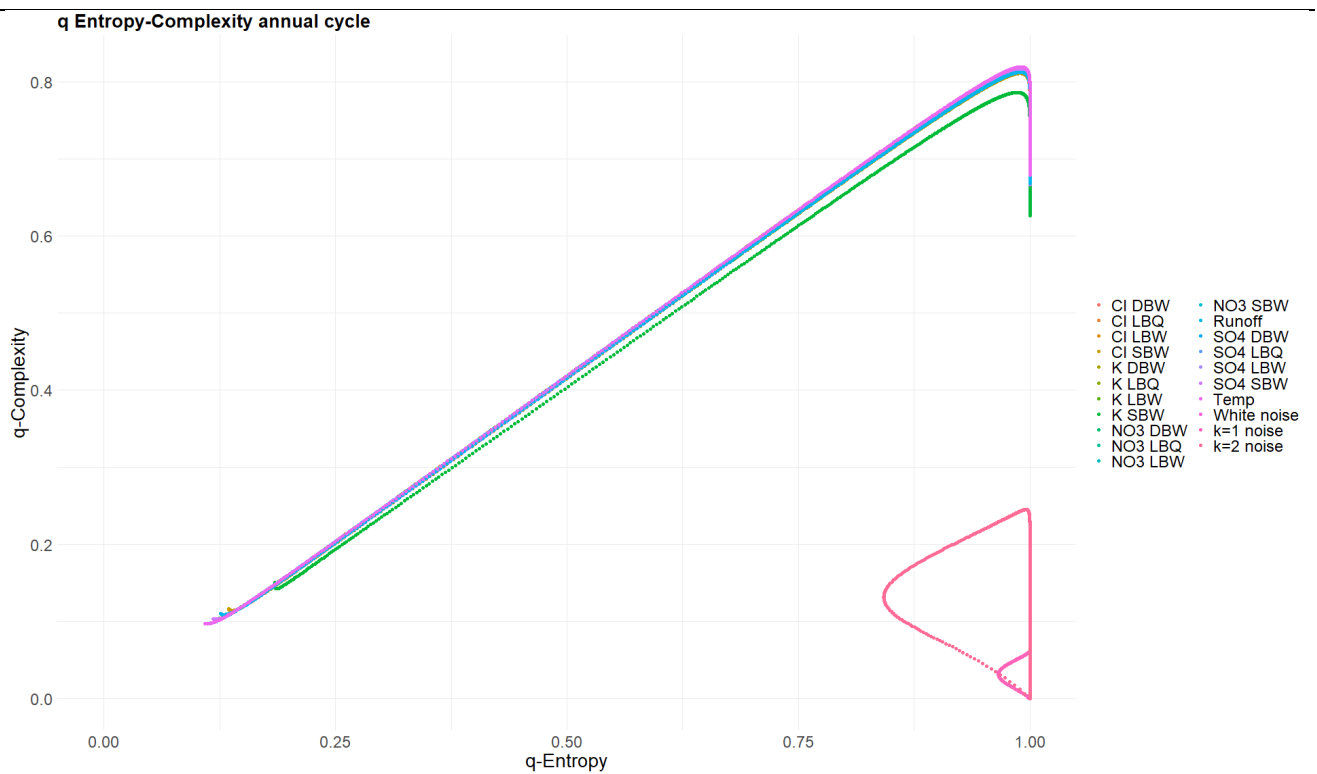

**Figure 17**: Same as Fig. 16, but for the annual components of the time series. The powernoise curves occupy a quite different region of the plane.

*3.5. Tarnopolski diagram*

The representation of a time series in the Abbe value/Turning points diagram is parameter free and does not use ordinal pattern statistics. Time series values are taken at face value without preprocessing. For some stochastic reference processes, the inventor of the method derived analytical results depending on time series length [31] and indicated regions in that plane occupied by time series models like ARMA(p,q) processes. The equation refer to the mean value of $\mathcal{A}$ and $T$ when generating a large set of similar time series of the given length; however, the finite size effects are already quite small for our $N = 860$.

Fig. 18 shows $\mathcal{A}$ and $T$ for all time series in all variants, together with the curves for fractional Brownian motion, fractional Gaussian noise and k-noise (the latter being a numerical result). The trend and annual components have zero or very low $\mathcal{A}$, quite contrary to the original or the detrended variants. The latter don't fit nicely to any of the three processes, cover a large range of $\mathcal{A}$ values and have $T$ values between fBm and k noise. These points certainly do not represent a one-dimensional curve, and won't fit to any simple stochastic process. In that regard, the Tarnopolski diagram is a rather different perspective on the time series compared to the entropy-complexity diagram.

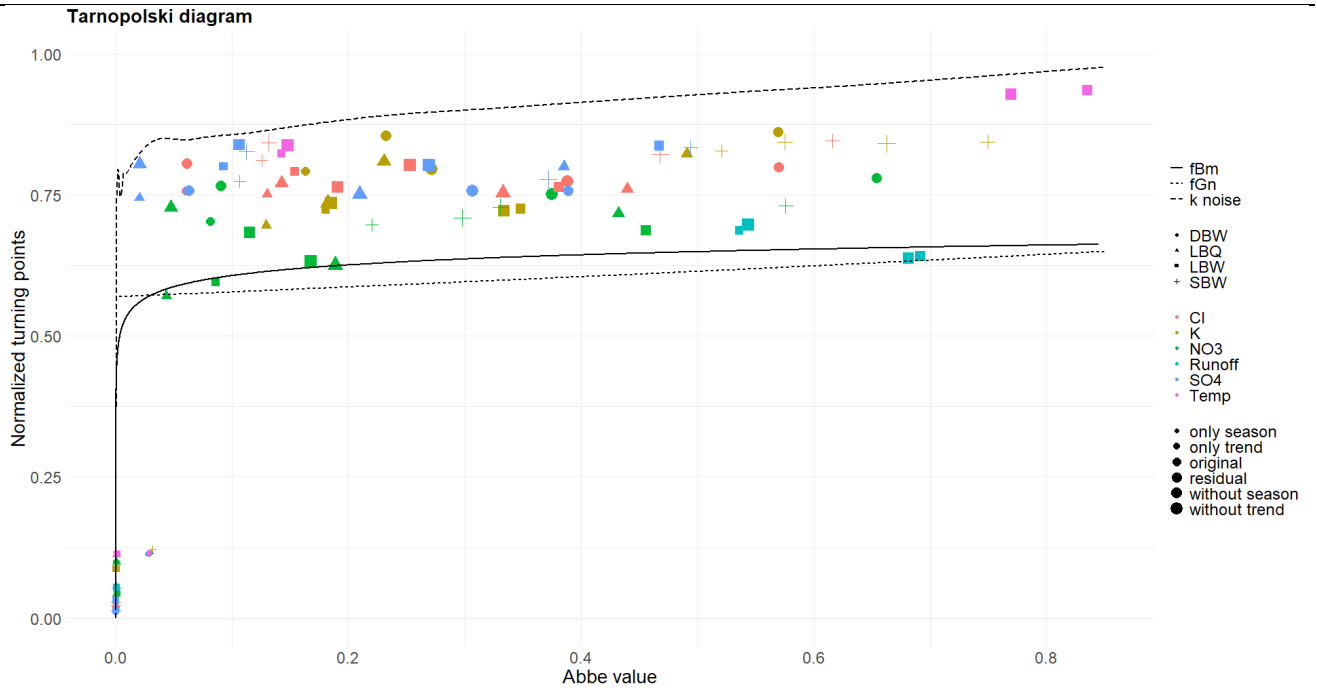

**Figure 18**: Tarnopolski diagram for the time series together with three different reference processes. The turning points are normalized to cover strictly the interval [0,1] ; the Abbe value extends to 1.5 which is reached by the fBM but not by our data sets, and is therefore not shown here.

*3.6. Horizontal Visibility Graph analysis*

The construction of the network of visibilities, considering every value of a time series as a node and two nodes connected with a link when the older one can "see" the later one in horizontal direction, is another nonparametric way to characterize the dynamical structure of a time series. From the networks obtained, we extract just one quantity: the slope of the decay of the degree distribution, assuming the exponential relationship of eq. (5). This assumption is empirically justified for many time series; however, at very large degrees, it always fails since there are gaps in the distribution - degrees which are simply non-existent in the distribution. We used a cutoff for the survival function of the degree distribution adapted to our time series length.

The slopes $\lambda_{HVG}$ for the exponential decay for the Horizontal Visibility Graphs obtained from the time series in this way are shown in Fig. 19. Here, the only reference process with a known theoretical value for this slope is white noise where $\lambda_{HVG}^{WN} = ln(3/2)$ ; this is drawn as a horizontal line in the Figure. For colored noise, it is not even known whether the degree distribution is of exponential type; and even assuming it, the spread of estimates for $\lambda_{HVG}$ for relatively short time series as is the case here is substantial [7], so we don't show them here. However, if anything, the slope values would be *larger* than for the white noise case.

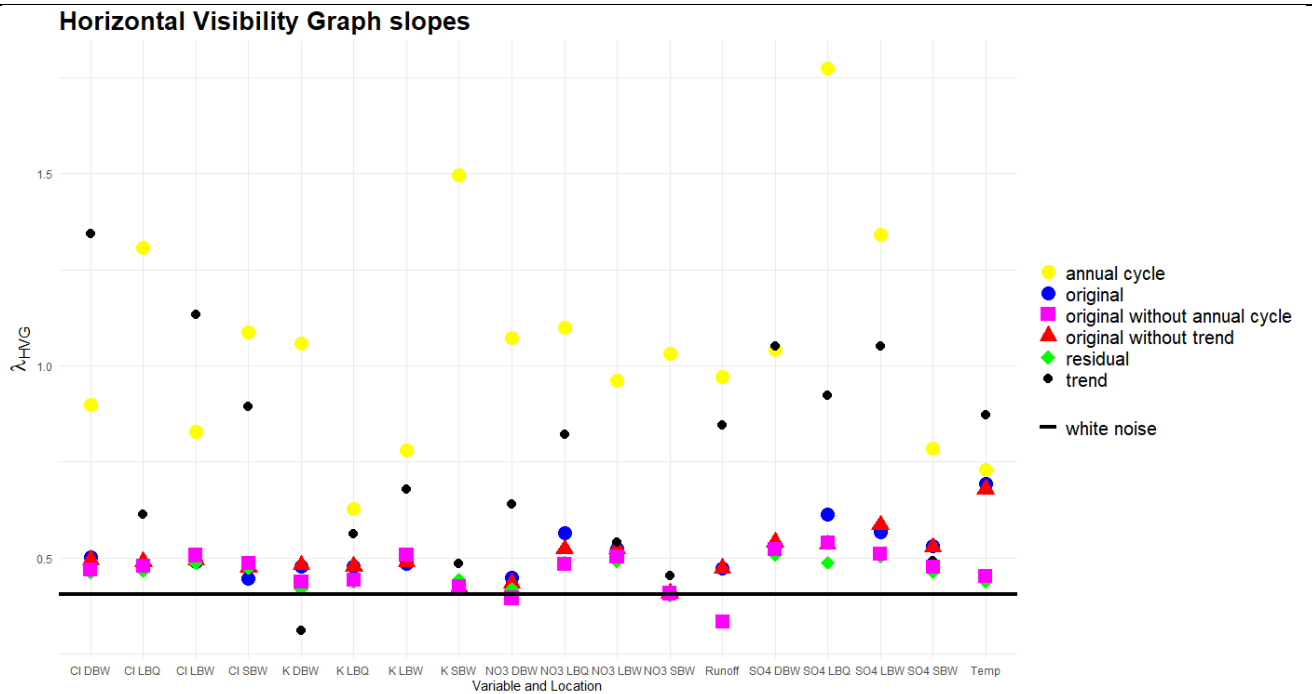

**Figure** 19: Slopes of the decay of the degree distribution of Horizontal Visibility Graphs

The HVG slope is rather insensitive to the removal of a trend component, and the original, detrended and the residual time series are by and large close to the white noise case from this perspective. However, a notable exception is temperature, and to a lesser extent also $NO_3$ and $SO_4$ at LBQ. The annual and trend component, however, extend to much higher slopes, their networks thus have a much tighter degree distribution, as is also reflected in smaller mean degrees for them (not shown). Notable outliers for the annual cycle are K at SBW (with a particular small annual amplitude, explaining only 4.5 % of the total variance), and $SO_4$ both at LBQ and LBW.

*3.5 COPPS analysis*

When applying the COPPS procedure to measured time series, the choices for the network depth $s$ and the embedding dimension $D$ are strongly constrained by the length of the time series, $N$: the number of possible groups/patterns is $\Lambda_s(N) = ((N-1)^2 + 1)^{s+1}$ [34]. The resulting complexity $\lambda_s(D)$ is also strongly dependent on $N$.

With our time series length $N = 860$, we chose $s = 1$ and $D = 4$. Fig. 20 shows the resulting $\lambda_1(4)$ for all variants. This is compared to results for Gaussian white noise, pink noise (k=1) and red noise (k=2), and the logistic map at fully developed chaos (r=4), all of which generated with the same length $N = 860$. The results for these processes are rather robust, repeating to generate time series of the same length and calculating $\lambda$ leads to rather narrow distributions (not shown).

Removing the trend or the annual component is always increasing the complexity. The magnitude of the effect is related to the strength (explained variance) of the component as expected; two extreme cases are the temperature where removing the annual cycle let the $\lambda$ parameter jump to almost the white noise case; and K LBWm where removal of the trend does not change $\lambda$ at all; the trend component for this ion has an explained variance of only 0.76% (Tab. 1). The complexity of the residuals is the highest in all cases. The opposite is true for the $\lambda$ values of the trend and annual component alone; they all are rather low, well below the reference value for red noise, with the trend $\lambda$ values being generally the lowest.

It would be possible to calculating an effective powernoise-$k$ for each time series, i.e. the one where the $\lambda$ value of the noise coincides with that of the observed time series.

However, the relative position of the noise and the deterministic chaotic map is not stable as a function of the time series length; at higher *N*, the noise curves cross the logistic map, all converging to a value of 1 for very long series, whereas the logistic map has a limit value well below that. Thus, conclusions on which "color" our time series have (which noise process they are resembling) depend on their length.

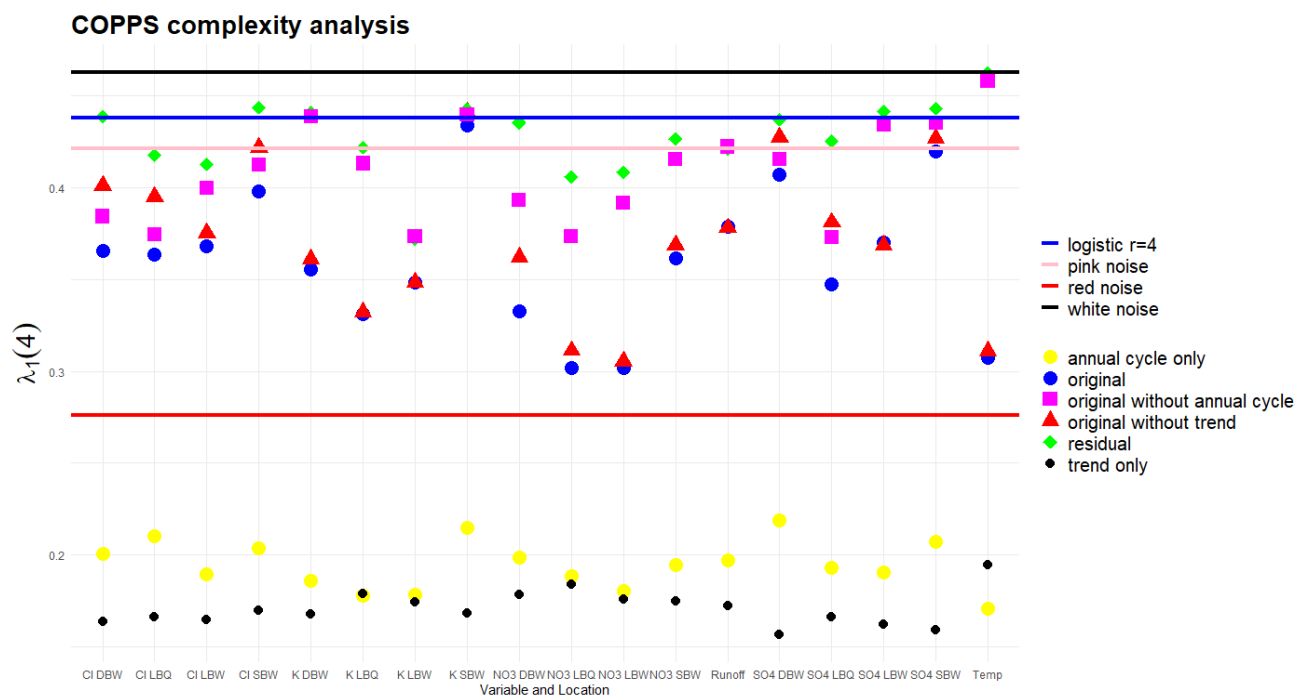

**Figure** 20: The complexity parameter $\lambda_{s=1}(D = 4)$ of the COPPS analysis [34]. Four reference stochastic processes are shown for comparison.

## 4. Discussion

The complexity analysis of the time series from the Bramke valley, and, we would claim, also for similar observations of water chemistry from other catchments, reveals that the stochastic component is overwhelmingly determining the dynamics of the system. From a linear perspective, this is surprising since the two components trends and seasonal cycle explain a lot of the total variance, in a few cases more than 90% (Table 1). However, the nonlinear properties of the time series dominate the complexity metrics considered.

Permutation Entropy and Complexity, Fisher Information and in particular Rényi and Tsallis entropy and complexity classify trend and seasonal cycle as deterministic signals. Although we chose a linear decomposition technique, SSA, for disentangling the deterministic from the stochastic part, the metrics show strong non-additivity; some quantifiers are almost unaffected by removal of one or even two of the deterministic signals, thus the residual is still a rather complex signal; it is in this sense that the stochastic parts of the time series are rather strong.

*4.1 Trend and Seasonality*

The ecosystems at the Bramke valley are exposed to a number of important environmental trends and disturbances. The major concern of the 1970ies and 1980ies, acidification, basically came to halt when flue-gas desulfurization was set into action. The consequence is a strong decline in the $SO_4$ concentrations in the streamwater of all four locations. Since this change is spatially large-scale, it is no surprise that the trend components for $SO_4$ are strongly correlated to each other (Figure A.2). $SO_4$ at DBW and SBW is also strongly correlated to temperature ($r = 0.71$ and $r = 0.88$, respectively), whereas at LBW and LBQ, $SO_4$ is anti-correlated ($r = -0.77$ and $r = -0.85$, respectively) with a stable phase shift of ca. 5 months (Fig. A.5). The annual cycle at LB reflects the expositional difference between a dry south-facing slope and a wet north-facing slope with $SO_4$ peaks in late winter. At the steeper, parallel sloping catchments DB and SB, the annual variation reflects probably an $SO_4$ depth gradient in soil storage.

For Cl, there is no linear trend but there is a decadal-long structure (Figure 6) which is also simultaneously present at the four locations, thus the correlation coefficients for the trend component are also very high in the Cl group (Figure A.2). The Cl found in the streams is to a large part from sea salt spray, which is also acting on broad spatial scales.

For $NO_3$ and K, the trends are more diverse between the four locations; for K, the strength of the trend component and also the amplitude of the annual cycle varies a lot between locations. At SBW, which was limed in 1989 and is the only catchment with a feasible areal cover of deciduous trees, $NO_3$ concentrations are much higher for most of the period compared to the other three ones; and the K dynamics is quite different from all other ions, as revealed already by the correlogram.

The residuals still show long-range correlations; it is unlikely that these are induced by periodic components at longer timescales as those were not discovered by the SSA.

*4.2 Complexity*

Trend and annual components show low entropy values and a MPR complexity higher than that of k noise at the same entropy level, again confirming their deterministic nature. The other variants are compatible with *k* noise, i.e. follow that curve, varying in their equivalent *k* value and thus in correlations strength; $NO_3$ appears to be the most complex; at the other end of the complexity spectrum, temperature residuals are almost white noise.

Contrary to pure long-term noise processes, all our time series contain missing patterns. For the stochastic variants, the number of missing patterns varies between 1 and 10 (out of 120). The most complex variable, $NO_3$, also has the highest number of missing patterns (Figure A.6). There is still the possibility that the missing patterns would occur for longer time series with the same dynamics [35]. For the deterministic parts, this number is substantially higher as expected, around 100 (not shown).

The Fisher Information for the stochastic variants (Fig. 13) is rather clumped together, but in all cases higher than for the *k* noise.

The q- and $\alpha$-entropies and complexities confirm the huge difference between the stochastic and the deterministic parts of our time series. It is difficult, however, to draw conclusions on differences within the stochastic group. As the area covered by the (almost) loops for the q-entropy-complexity relation depends on the Hurst parameter for fBM, one could calculate an effective Hurst parameter for our time series and compare to the one obtained with more conventional methods.

*4.3 Tarnopolski diagram*

The Abbe value of the Tarnopolski diagram might be used as a classifier for the dynamics of our time series as it spreads over almost the whole available range. They do not seem to make a distinction between the locations easy; the more complex variables (as judged from the MPR plot and Fisher Information plot) appear at lower Abbe values. For the stochastic part, the number of turning points is generally higher and the position of our time series interpolates between fBm and $k$ noise.

*4.3 HVG slopes*

Among the quantifiers considered, this might be the least discriminating for the stochastic parts. Most of the $\lambda_{HVG}$ values obtained are rather close to white noise, and removing trend or seasonality has a minor effect. The slopes for the deterministic parts are very different, so this basic distinction is possible also here.

*4.4 COPPS*

The slopes for the ordinal patterns interpolate between red and pink noise in most cases for the stochastic parts. The effect of removing trend or seasonality differs between the variables and is related to the strength of these components (explained variance percentages). The slopes are increasing from the original to the residual, and reach values compatible with the logistic map e.g. for $SO_4$. The location SBW reaches the highest slopes for the residuals in three of four cases. Trend and seasonal component are characterized by rather low $\lambda$ values. The COPPS slope seems to be an alternative good indicator for complexity of a time series; however, the strong dependence on time series length necessitates comparisons only between time series of the same length.

*4.5 Summary*

Complexity and information measures were used here as efficient tools comparing and classifying this set of environmental time series. They separate the classification task, e.g. between stochastic and deterministic parts, from the more difficult issue of reconstructing a given time series. We find similarities in dissolved ions across sites (especially Cl), but also highly site-specific behavior (K at SBW). Especially for the behavior of variables for which biotic interactions are implicated (K, $NO_3$, water), reference processes have been difficult to identify. Some of the time series are easy to classify by these methods, but difficult to reproduce (and explain) by the respective process-based models. The suite of methods presented here is a rather stringent test environment for them.

Complexity measures have become a critical tool to compare documented environmental behavior relative to candidate references processes from various disciplines. They may thus form a building in search for a common formalization of environmental processes.

**Supplementary Materials:** Supporting Information will be added prior to publication.

**Author Contributions:** Conceptualization, H.L. and M.H.; methodology, H.L.; software, H.L. and M.H.; validation, H.L. and M.H.; formal analysis, H.L.; investigation, H.L.; resources, M.H.; data curation, M.H.; writing—original draft preparation, H.L.; writing—review and editing, M.H.; visualization, H.L.; funding acquisition, H.L. All authors have read and agreed to the published version of the manuscript."

**Funding:** H.L. acknowledges partial financial support from the Research Council of Norway (Contract No. 342631/L10) and the EU H2020 Climb-Forest project (101059888).

**Institutional Review Board Statement:** Not applicable.

**Data Availability Statement:** The time series of the Bramke catchments will be made available prior to publication.

**Acknowledgements:** It is our great pleasure to thank Prof. Osvaldo Rosso for collaboration during the last 15 years, and his hospitality during the sabbatical stay of the first author of this contribution in Maceió. Osvaldo's ideas and method development were the main inspiration for this study. The data used here are collected, shared and discussed with us by Dr. Henning Meesenburg (Northwest German Forest Research Institute at Göttingen, Germany) since many years. The work here could not be done without his permanent support. Katharina Funk and Britta Aufgebauer (Univ. of Bayreuth, Germany) helped organizing the data set and contributed to software development. We also would like to thank Prof. Eyebe Fouda (University of Yaoundé, Cameroon) for sharing his COPPS code with us and fruitful discussions. The complexity measures were calculated using the R package **statcomp** which Sebastian Sippel (now Univ. of Leipzig, Germany) developed, and the phase shift calculations used the package **simsalabim** which Lukas Gudmundsson (now ETH Zürich, Switzerland) developed, both in the context of their respective master theses where the two authors of this article were their supervisors.

**Conflicts of Interest:** The authors declare no conflicts of interest. Also, the funders had no role in the design of the study; in the collection, analyses, or interpretation of data; in the writing of the manuscript; or in the decision to publish the results.

## Appendix A

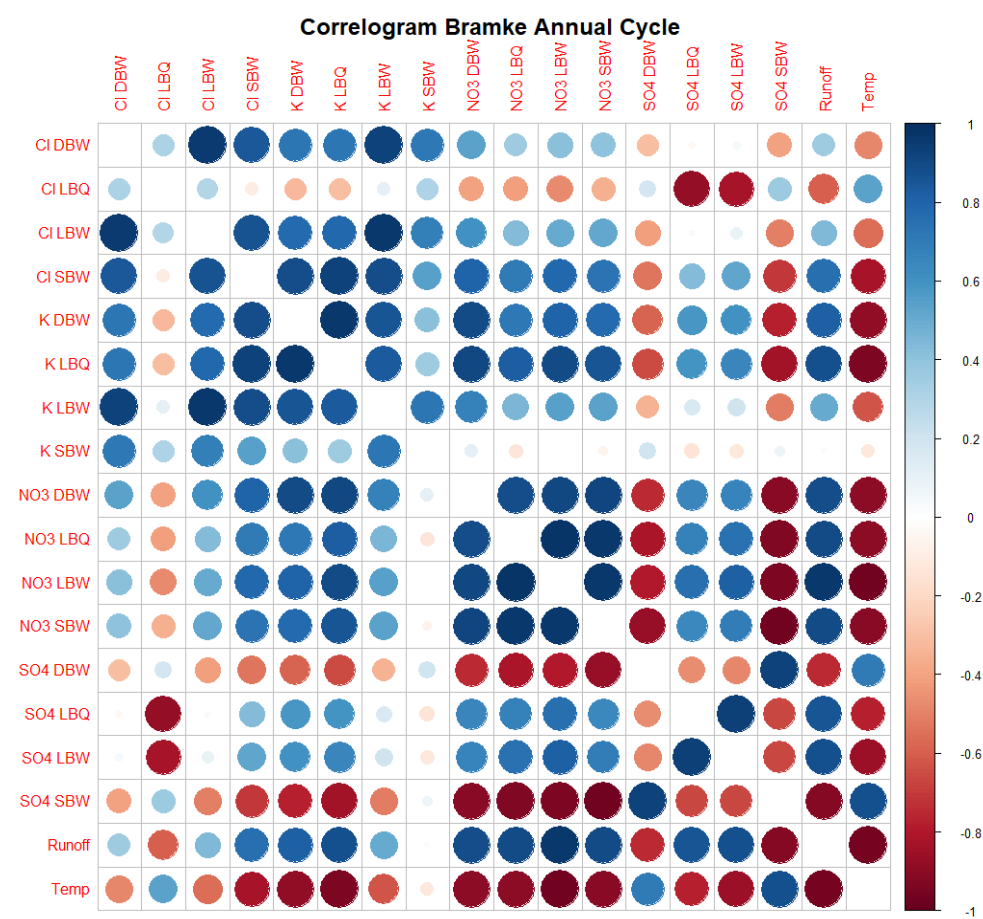

**Figure A.1:** Correlogram for the annual cycle alone.

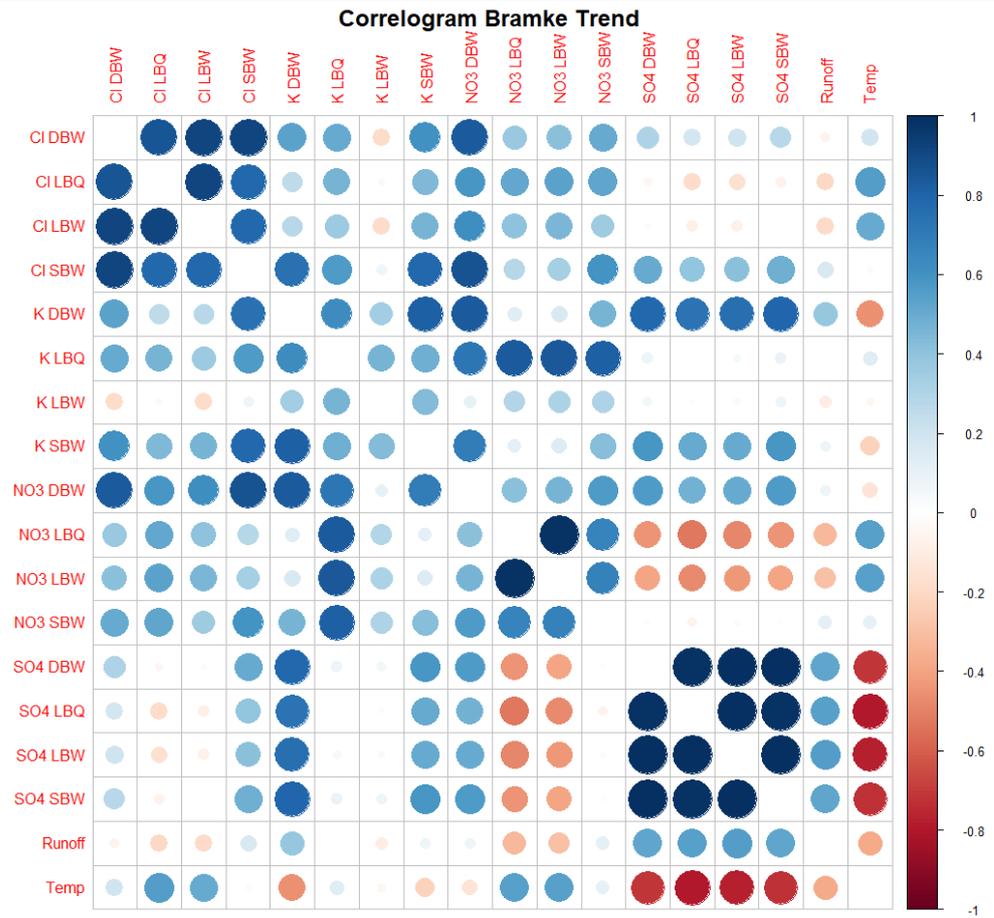

**Figure A.2:** Correlogram for the trend component alone.

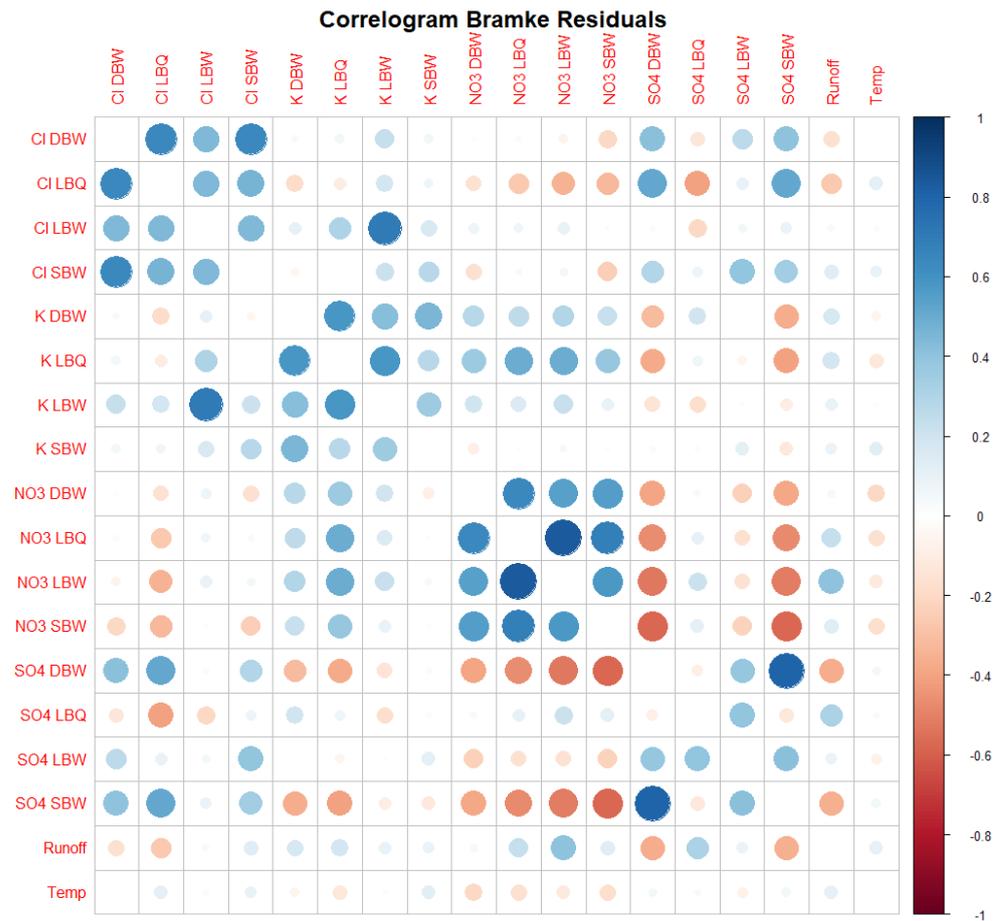

**Figure A.3:** Correlogram for the residuals (after removing both the trend and the annual component)

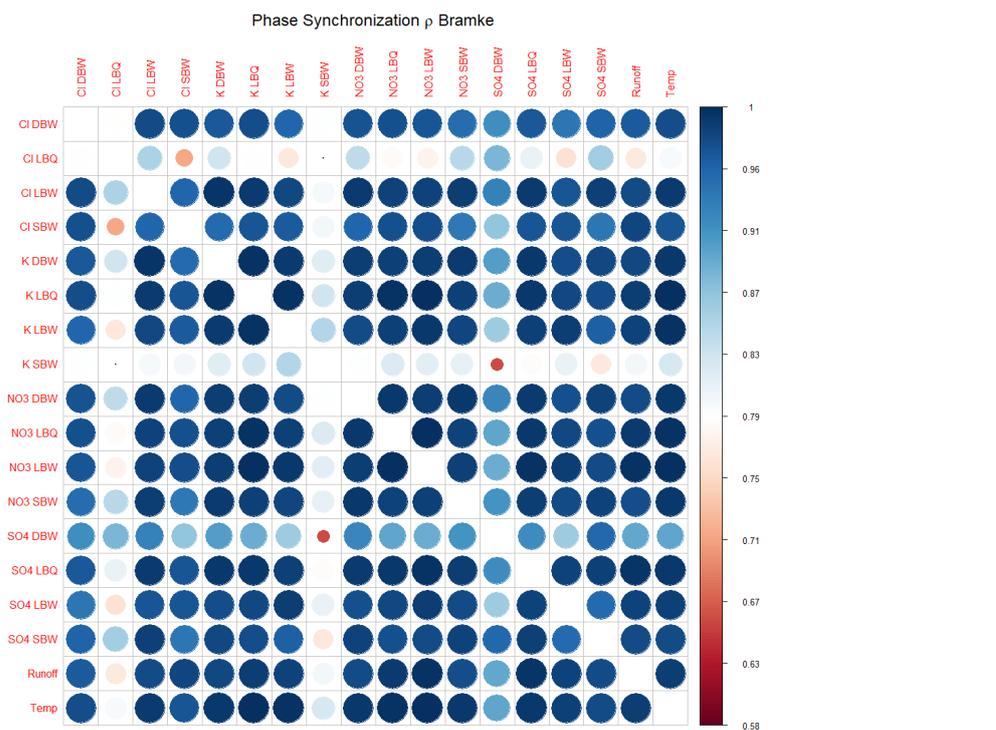

**Figure A.4:** Phase synchronization index $\rho$ for the annual component of the time series.

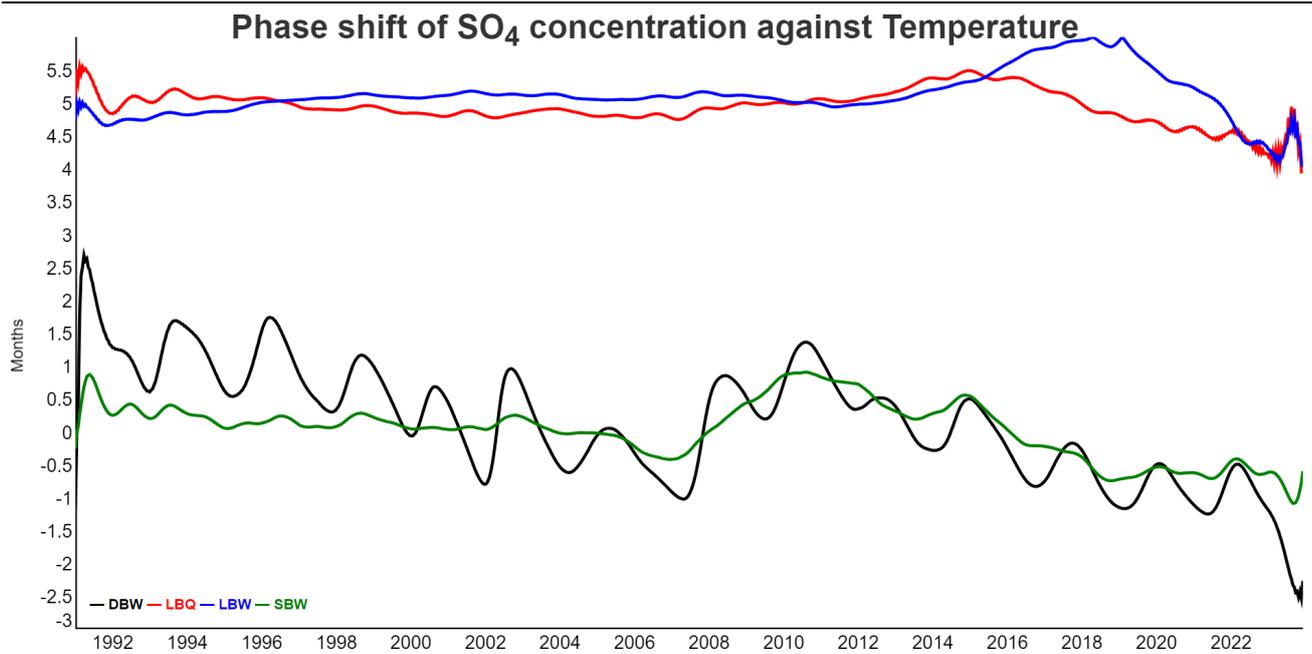

**Figure A.5**: Phase shift time series for SO$_4$ at the four locations and temperature

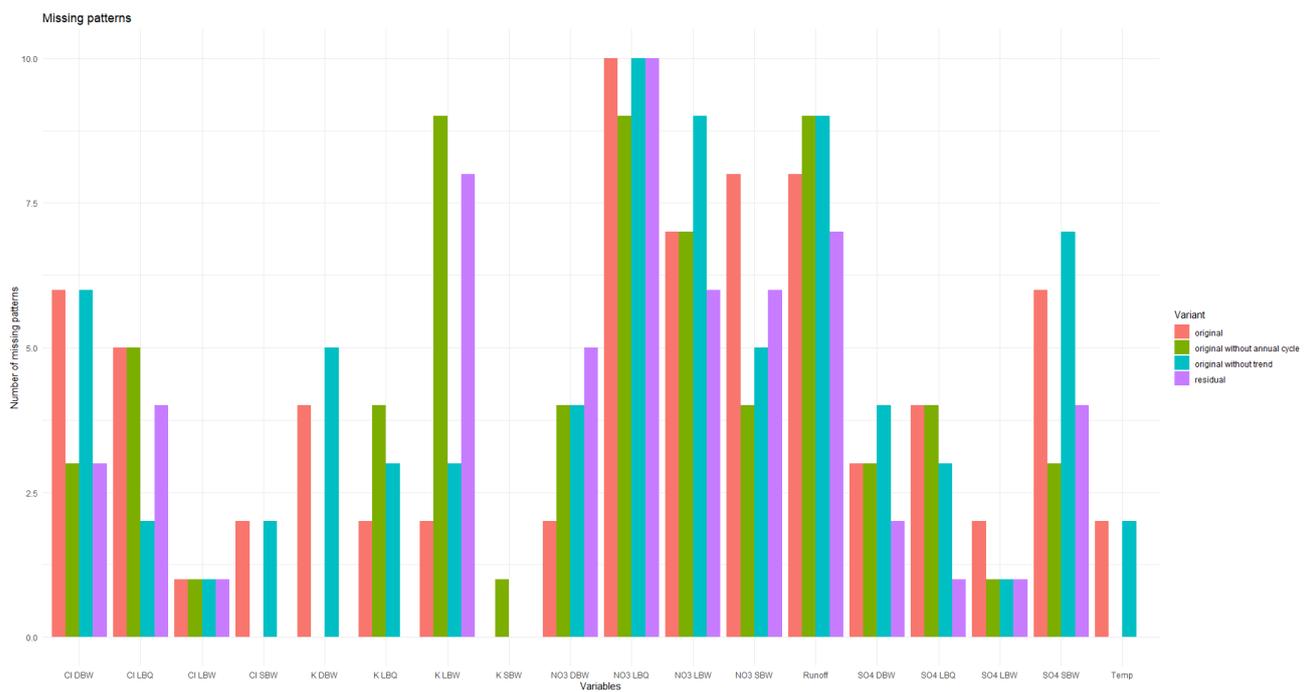

**Figure A.6:** The number of missing patterns for the original, detrended, deseasonalized, and residual variant.